\documentclass[twocolumn,prb,aps,nobibnotes]{revtex4}
\usepackage{graphicx}%
\usepackage{dcolumn}
\usepackage{amsmath}   
\usepackage{bm}
\usepackage{pstricks,pst-node,pst-plot}
\usepackage{longtable}
\begin{document}

\title{\centering\Large\bf Redox Entropy of Plastocyanin: 
                    Developing a Microscopic View of Mesoscopic
                    Polar Solvation }
\author{David N. LeBard}
\author{Dmitry V.\ Matyushov}
\affiliation{Center for Biological Physics, Arizona State University, 
             PO Box 871604, Tempe, AZ 85287-1604}
\date{\today }
\begin{abstract}
  We report applications of analytical formalisms and Molecular
  Dynamics (MD) simulations to the calculation of redox entropy of
  plastocyanin metalloprotein in aqueous solution. The goal of our
  analysis is to establish critical components of the theory required
  to describe polar solvation at the mesoscopic scale. The analytical
  techniques include a microscopic formalism based on structure
  factors of the solvent dipolar orientations and density and
  continuum dielectric theories. The microscopic theory employs the
  atomistic structure of the protein with force-field atomic charges
  and solvent structure factors obtained from separate MD simulations
  of the homogeneous solvent. The MD simulations provide linear
  response solvation free energies and reorganization energies of
  electron transfer in the temperature range 280--310 K. We found that
  continuum models universally underestimate solvation entropies, and
  a more favorable agreement is reported between the microscopic
  calculations and MD simulations. The analysis of simulations also
  suggests that difficulties of extending standard formalisms to
  protein solvation are related to the inhomogeneous structure of the
  solvation shell at the protein-water interface combining
  islands of highly structured water around ionized residues along
  with partial dewetting of hydrophobic patches.  Quantitative
  theories of electrostatic protein hydration need to incorporate
  realistic density profile of water at the protein-water interface.
\end{abstract}

\preprint{Submitted to \textit{J.\ Chem.\ Phys.} }
\maketitle

\section{Introduction}
\label{sec:1}
Calculations of the thermodynamics of hydrated biopolymers present a
significant challenge to theoretical algorithms. In many cases the
problem can be treated by numerical simulations with force fields
assigned to the biomolecule (solute) and water
(solvent).\cite{Brooks:88} The obvious difficulty is the large
computational load and still existing uncertainties in the treatment
of the long-range electrostatics.  The problem, however, becomes more
nontrivial when derivatives of thermodynamic potentials, e.\ g.\ redox
entropy, need to be computed or when the solvation thermodynamics
changes on the time- and length-scale unattainable to standard
Molecular Dynamics (MD) protocols, e.\ g.\ in problems related to
protein folding.\cite{Eisenberg:86,Zhou:04} In all such cases, coarse
graining of the system is required and that can be done on various
length-scales.\cite{Mueller:06,Gohlke:06} Dielectric continuum
algorithms, solving the boundary Poisson problem, are computationally
very efficient. In this approximation, all length-scales below the
largest distance of microscopic correlations (density and/or
polarization) are averaged out into a continuum surrounding the
solute. These approaches are normally represented by either direct
solution of the Poisson-Boltzmann equation on the real-space
grid\cite{delphi02} or even more approximate formalisms under the
umbrella of the generalized Born approximation.\cite{Schaefer:96}

When the cavity cut out by the solute from the continuum dielectric is
properly parametrized, equations of continuum electrostatics provide a
reasonable estimate of the solvation Gibbs
energy.\cite{Gunner:91,Sharp:98,Siriwong:03} The fundamental problem
of this approach is that the local structure of the solvent around the
solute, averaged out in the continuum limit, is what effectively forms
the dielectric cavity.  While this structure can be parametrized by
choosing proper van der Waals (vdW) radii,\cite{Qiu:97} this
parametrization needs to be re-done every time the thermodynamic state
of the solvent changes.  This difficulty makes continuum formalisms
unreliable for the calculation of derivatives of the Gibbs energy, for
instance the entropy of
solvation.\cite{DMjpcb:99,DMjcp2:04,DMcp:06,DMjpca1:06} In addition,
the surface of a protein is highly heterogeneous combining hydrophobic
patches exposing non-polar residues and hydrophilic patches made of
ionized/polar residues. While the water structure is rigid around ionized
residues, probably resembling the well-studied case of solvation
around simple ions, water is much less structured at hydrophobic
patches with the potential for dewetting\cite{ZhouScience:04} or/and
oscillations of the water
occupation.\cite{Tarek_PhysRevLett:02,Choudhury:07} It is clear that
simplistic continuum does not represent this complex
reality,\cite{Zhou:04} and one needs to incorporate the ability of the
solvent to fluctuate into the solvation model.

The goal of this paper is to extend the microscopic view of solvation
in polar solvents, which we have been developing in the past in
application to small and medium-size
solutes,\cite{DMcp:93,DMjcp2:04,DMcp:06} to solvation of solutes of
mesoscopic dimension, biopolymers in the first place. The length-scale
of this problem presents an obvious obstacle to numerical simulation
techniques. On the other hand, the same length-scale allows one to
hope that some of the short-range features of the solvent structure
around the solute, making solvation of small molecules so specific,
might average out on a larger scale. If this averaging is realized for
solvation of biopolymers, it would allow coarse-grained models to
efficiently operate in this field complementing direct numerical
simulations. Our approach to the problem is to coarse-grain the
solvent response into a number of solvent correlation functions
(structure factors) representing the nuclear modes affecting
electrostatic solvation. The microscopic nature of the solvent
response is then incorporated into the wave-vector dependence of these
structure factors efficiently filtering out the length-scales
insignificant for solvation.

This study poses the central question for the future development of
such techniques: What are the solvent modes which play the central
role in the thermodynamics of mesoscopic polar solvation and what are
the theory ingredients critical for capturing the basic physics of
large-scale solvation? We study this problem here by carrying out
extensive MD simulations of solvation of plastocyanin (PC) in TIP3P
water in the temperature range 280--310 K. This fully atomistic
approach is compared to continuum electrostatics and to our
microscopic algorithm, operating with $\mathbf{k}$-space correlation
functions, which was designed to scale efficiently on the mesoscopic
length-scale.

\begin{figure}[htbp]
  \centering
  \includegraphics*[width=7cm]{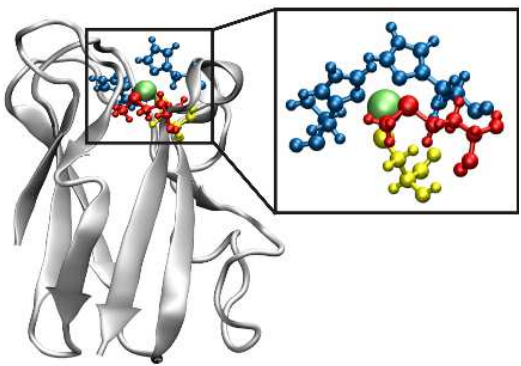}
  \caption{Structure of plastocyanin: the active site includes copper
    ion (green), 2 histidines (blue), methionine (red), and cysteine
    (orange) residues.}
  \label{fig:1}
\end{figure}

Plastocyanin from spinach is a single polypeptide chain of 99 residues
forming a $\beta$-sandwich, with a single copper ion coordinated by 2
sulfurs from cysteine and methionine and 2 nitrogens from histidine
residues (Fig.\ \ref{fig:1}). The presence of the copper ion, which
can change redox state, allows PC to function as a mobile electron
carrier in the photosynthetic apparatus of plants and bacteria. It
accepts an electron from ferrocytochrome \textit{f} and diffusionally
carries it to another docking location where the electron is donated
to the oxidized form of Photosystem I.\cite{Ubbink:98}

\begin{figure}
  \centering
  \includegraphics*[width=8cm]{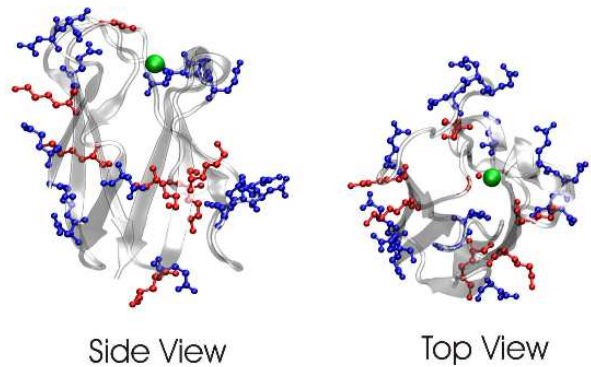}
  \caption{Distribution of the positive and negative charge on the
    surface of the protein. The positively and negatively charged
    residues are shown, respectively, in red and blue.  The copper ion
    is shown in green. }
  \label{fig:2}
\end{figure}

The redox thermodynamics of PC has been characterized
experimentally\cite{Sailasuta:79,Sola:99,Battistuzzi:03} and combined
quantum/simulation calculations have been done as
well.\cite{Lockwood:01,Datta:04} The early focus of the theoretical
studies had been on unusually high redox potentials of copper
proteins, which was assigned to the non-traditional distorted
tetrahedral coordination on the copper
ion.\cite{Hansen:04,Solomon:04,Solomon:06} In particular, the Cu-S
bond to methionine is unusually long and is actually broken in the
reduced state of PC at pH $<3.8$.\cite{Guss:86} The protein is also
highly charged at pH$\simeq 7$ ($-9.0$ in reduced state and $-8.0$ in
oxidized states). The charge is made by 15 negatively charged
deprotonated residues (9 glutamic and 6 aspartic acids) and six
positively charged lysine residues with amino groups protonated (Fig.\
\ref{fig:2}).  The asymmetric charge distribution located on the
protein surface creates the dipole moment of 2200 D in the oxidized
state (Ox) and of 2470 D in the reduced state (Red), both numbers are
calculated relative to the center of partial charges.

The redox potential of the protein includes a component from the local
ligand field of the active site and the Gibbs energy of solvation. The
computation of the former requires quantum mechanics, making the
problem of calculating the overall redox potential a very non-trivial
exercise.\cite{Stephens:96,Olsson:03,Datta:04} Calculations of
solvation thermodynamics can be reasonably accomplished using partial
atomic charges parametrized from quantum calculations in the vacuum.
The experimental input comes from measurements of redox
entropy\cite{Yee:79,Sailasuta:79,Sola:99} since the
temperature-independent ligand-field component is expected to vanish
in the temperature derivative.

\section{Microscopic Solvation Model}
\label{sec:2}
The principal idea of the microscopic solvation model is to reduce the
problem of solvation of an arbitrary solute in a polar solvent to a
formalism combining two major blocks: electrostatics of an isolated
solute and non-local correlation functions of the pure solvent. The
idea of assembling separate solute and solvent properties in a
solvation model is obviously not new going back to Born\cite{Born:20}
and Onsager\cite{Onsager:36} and all the subsequent development of
continuum electrostatics in application to
solvation.\cite{Kirkwood:34,Gunner:91,Schaefer:96,Sharp:98,delphi02}
The advantage of our approach is in avoiding the necessity to know the
microscopic solute-solvent structure, which is the main complexity of
microscopic solvation models and is also their main advantage when the
problem is successfully resolved by either solving integral
equations\cite{Raineri:99} or by applying
time-dependent\cite{Yoshimori:04,BagchiCP:06} or
equilibrium\cite{Biben:98,Ramirez:02,RamirezJPCB:05} density
functional methods. Inserting a solute into a dense liquid creates a
significant distortion of its structure, and the incomplete account of
the coupling between the short-range density profile around the solute
with the long-range polarization field is perhaps the weakest part of
our formulation when applied to small solutes.\cite{DMjcp3:06} [This
deficiency is almost completely off-set by averaging of the density
profile around a nano-scale solute (see below).] On the other hand, a
strong side of our formalism, its ability to treat solvation of large
solutes of irregular shape and arbitrary charge
distribution,\cite{DMjpcb1:03,DMcp:06} becomes particularly useful in
application to protein solvation.

The reduction of the many-body solvation problem to an irreducible
representation in terms of a few basic correlation functions depends
on the symmetry of the solute-solvent interaction potential. The
number of correlation functions is known to grow with increasing the
rank of the solvent multipole included in the interaction
potential.\cite{DMjcp4:05} Solvent dipoles are for the most part
sufficient for solvation in polar liquids,\cite{DMjpca:01} in which
case the solute-solvent interaction potential $V_{0s}$ (``0'' and
``s'' are used for the solute and solvent, respectively) is a sum of
pairwise interactions of the solute electric field
$\mathbf{E}_0(\mathbf{r})$ with the solvent dipoles
\begin{equation}
  \label{eq:1-1}
  V_{0s} = - \sum_{j=1}^N \mathbf{m}_j'\cdot \mathbf{E}_0(\mathbf{r}_j)  .
\end{equation}
Here, $\mathbf{m}_j'$ is the dipole moment characteristic of the bulk
state of the solvent;  $m'$ is usually higher that the vacuum dipole
$m$ because of the collective field of the induced solvent
dipoles.\cite{SPH:81} For instance, the dipole moment of water in the
liquid state, 2.4--2.6 D,\cite{CReynolds:01} is higher than the
gas-phase dipole of 1.83 D.

We will focus on the electrostatic component of the chemical potential
of solvation $\mu_{0s}$ which contains all the information relevant to
electrostatic solvation. Linear response approximation (LRA)
significantly simplifies the problem and provides several equivalent
routes to $\mu_{0s}$. One can consider the full interaction between the
atomic charges of the solute and the solvent and determine $\mu_{0s}$ as
half of the average solute-solvent interaction energy,\cite{DMjcp1:99}
$\mu_{0s}=\langle V_{0s}\rangle/2 $.  Alternatively, one can use the second
cumulants, $\langle(\delta V_{0s})^2\rangle$ or $\langle(\delta V_{0s})^2\rangle_0$.\cite{DMjcp2:04} In
the first cumulant, the angular brackets $\langle \dots \rangle$ refer to an
ensemble average over the fluctuations $\delta V_{0s}$ in the solvent in
equilibrium with the full charge distribution of the solute. For the
second cumulant, $\langle \dots \rangle_0$ implies that all the charges of the
solute have been set to zero, and fluctuations of the solvent in the
solute vicinity are regulated only by short-range solute-solvent
interactions, molecular repulsions in the first place.  In the LRA,
the two cumulants are equal,\cite{Carter:91} which physically means
that the solute electrostatic forces do not significantly change the
solvent structure around the solute established by the prevalence of
short-range repulsions.\cite{WCA} Computer simulations for the
most part support this
picture\cite{Hwang:87,Kuharski:88,Blumberger:05} with a few exceptions
of very strong solute-solvent electrostatic coupling found for small
solutes.\cite{Carter:91,Fonseca:94,DMjpca1:06}

This observation opens up a significant simplification of the
calculation algorithms. Instead of solving the inhomogeneous problem of
restructuring the solvent in an external field of the solute, it
appears to be sufficient to look at the statistics of solvent
fluctuations around the repulsive core of the solute. This strategy is
used here and we will base our calculations on the relation
\begin{equation}
  \label{eq:1-2}
  -\mu_{0s}= (\beta/2) \langle(\delta V_{0s})^2\rangle_0 ,
\end{equation}
where $\delta V_{0s}= V_{0s} - \langle V_{0s}\rangle_0$ and $\beta=1/(k_{\text{B}}T)$. 

By using the interaction potential according to Eq.\ \eqref{eq:1-1}, one
can re-write Eq.\ \eqref{eq:1-2} in the form typical for Gaussian (LRA)
models of solvation \cite{Lee:88,Chandler:93}
\begin{equation}
  \label{eq:1-3}
  -\mu_{0s} = \frac{1}{2} 
   \mathbf{\tilde E}_0(\mathbf{k}_1)* \bm{\chi}(\mathbf{k}_1,\mathbf{k}_2)*
     \mathbf{\tilde E}_0(\mathbf{k}_2) . 
\end{equation}
Here, the 2-rank tensor $\bm{\chi}(\mathbf{k}_1,\mathbf{k}_2)$ is the
response function\cite{Hansen:03} of the system composed of a dipolar
solvent and a solute to a weak field of the solute. The inhomogeneous
character of the problem is reflected by the fact that
$\bm{\chi}(\mathbf{k}_1,\mathbf{k}_2)$ depends on two wave-vectors,
$\mathbf{k}_1$ and $\mathbf{k}_2$, separately and not on
$\mathbf{k}_1-\mathbf{k}_2$, as is the case with response functions of
homogeneous solvents.  The asterisk in Eq.\ \eqref{eq:1-3} refers to
both the tensor contraction and integration in inverted
$\mathbf{k}$-space. In addition, $\mathbf{\tilde E}_0(\mathbf{k})$ is
the Fourier transform of the electric field of the solute defined by
the integral limited to the solvent volume $\Omega$:
\begin{equation}
  \label{eq:1-4}
  \mathbf{\tilde E}_0(\mathbf{k}) 
      = \int_{\Omega} \mathbf{E}_0(\mathbf{r}) e^{i\mathbf{k}\cdot\mathbf{r}} d\mathbf{r} .
\end{equation}
The shape of the solute thus enters  both the response function
$\bm{\chi}(\mathbf{k}_1,\mathbf{k}_2)$ and the field Fourier transform
$\mathbf{\tilde E}_0(\mathbf{k})$.  The charge distribution of the
solute, which determines the electric field $\mathbf{\tilde
  E}_0(\mathbf{k})$, is given by its electronic density and is
commonly represented by partial atomic charges.

The main challenge of this formalism, as well as of other Gaussian
solvation theories,\cite{Weeks:02} is how to connect the inhomogeneous
response function $\bm{\chi}(\mathbf{k}_1,\mathbf{k}_2)$ to the shape of
the solute repulsive core and the self-correlation functions of the
solvent modes affecting solvation. Two modes naturally appear in most
theories: dipolar (orientational) polarization and density
fluctuations.\cite{Fried:90,Bagchi:91,DMcp:93,Ramirez:02} For the
former, the combination of axial symmetry introduced by the
wave-vector $\mathbf{k}$ with the vector character of the dipolar
polarization $\mathbf{P}(\mathbf{k})$ allows one to split the 2-rank
tensor $\bm{\chi}_s(\mathbf{k})=(\beta/ \Omega)\langle|\delta \mathbf{P}(\mathbf{k})|^2\rangle$
into the longitudinal and transverse dyads:\cite{Madden:84}
\begin{equation}
  \label{eq:1-5}
  \bm{\chi}_s(\mathbf{k}) = \frac{3y}{4\pi}\left[\mathbf{J}^LS^L(k) + 
      \mathbf{J}^T S^T(k)\right] ,
\end{equation}
where $\mathbf{J}^L=\mathbf{\hat k}\mathbf{\hat k}$,
$\mathbf{J}^T=\mathbf{1} - \mathbf{\hat k}\mathbf{\hat k}$. In Eq.\
\eqref{eq:1-5}, $y$ is the effective density of both permanent and
induced dipoles in the liquid which commonly appears in theories of
dielectrics\cite{SPH:81}
\begin{equation}
  \label{eq:1-6}
  y = (4\pi /9) \beta m'^2 \rho + (4\pi/3)\rho\alpha.
\end{equation}
In Eq.\ (\ref{eq:1-6}), $\alpha$ is the dipolar polarizability of the
solvent particle.

The scalar functions $S^L(k)$ and $S^T(k)$ in Eq.\ \eqref{eq:1-4} are,
correspondingly, the longitudinal and transverse structure factors of
dipolar fluctuations of the homogeneous solvent (see below).  The
$k=0$ values of these structure factors are related to the dielectric
constant, $\epsilon_s$, by the following equations:
\begin{equation}
  \label{eq:1-14-1}
  \begin{split}
    S^L(0) & = (\epsilon_s -1) /(3y\epsilon_s), \\
    S^T(0) & = (\epsilon_s -1) /(3y).
  \end{split}
\end{equation}
Also, the trace of $\bm{\chi}_s(0)$ over the Cartesian projections is
proportional to the Kirkwood $g$-factor\cite{Boettcher:73}
\begin{equation}
  \label{eq:1-14}
  g_{\text{K}} = \frac{1}{3} \left[ S^L(0) + 2 S^T(0)\right] .
\end{equation}

The expansion of the solvation chemical potential in the Mayer
functions corresponding to the solute-solvent interaction potential
leads to the following form for the response
function\cite{DMcp:93,DMjcp2:04}
\begin{equation}
  \label{eq:1-6-1}
  \bm{\chi}(\mathbf{k}_1,\mathbf{k}_2) =
   \bm{\chi}_p(\mathbf{k}_1,\mathbf{k}_2) + \bm{\chi}_d(\mathbf{k}_1,\mathbf{k}_2) ,
\end{equation}
where
\begin{equation}
  \label{eq:1-6-2}
   \bm{\chi}_p(\mathbf{k}_1,\mathbf{k}_2)=\bm{\chi}_s(\mathbf{k}_1)\delta_{\mathbf{k}_1,\mathbf{k}_2} .
\end{equation}
In Eq.\ \eqref{eq:1-6-2}, $\Omega\delta_{\mathbf{k}_1,\mathbf{k}_2}=(2\pi)^3
\delta(\mathbf{k}_1-\mathbf{k}_2)$ is the Kronecker symbol and
$\bm{\chi}_d(\mathbf{k}_1,\mathbf{k}_2)$ in Eq.\ (\ref{eq:1-6-1}) is the
component of the response originating from the local fluctuations of
the solvent density around the solute\cite{DMcp:93}
\begin{equation}
  \label{eq:1-6-3}
  \bm{\chi}_d(\mathbf{k}_1,\mathbf{k}_2) 
          = (3y / 8\pi) (1 -
          S(k_1))\theta_0\left(\mathbf{k}_1-\mathbf{k}_2\right) .
\end{equation}
Here, $S(k)=N^{-1}\langle|\delta\rho(\mathbf{k})|^2\rangle$ is the density-density
structure factor of the homogeneous solvent and $N$ is the number of
solvent molecules. In addition, $\theta_0(\mathbf{k})$ is the Fourier
transform of the step function $\theta_0(\mathbf{r})$ defining the solute
shape. It is equal to unity for $\mathbf{r}$ inside the solute and is
zero otherwise.

The problem with the direct perturbation result in Eq.\
\eqref{eq:1-6-1} is that it contains the transverse polarization
response function $\propto S^T(k)$ diverging in its continuum, $k\to 0$, limit
as the solvent dielectric constant goes to infinity
[Eq.\ (\ref{eq:1-14-1})]. The problem is really caused by the
non-spherical shape of the solute. The electric field of the solute
charges is longitudinal. However, when the symmetry of the solute is
different from the symmetry of the charge distribution in a sense that
the cavity boundary does not coincide with the equipotential surface,
the Fourier integral in Eq.\ \eqref{eq:1-4} generates a transverse
component of $\mathbf{\tilde E}_0(\mathbf{k})$. Notice that this is
always the case when electron transfer reactions are
considered.\cite{KKV:76} A transverse component in $\mathbf{\tilde
  E}_0(\mathbf{k})$ then results in a ``transverse catastrophe'' for
solvents of high polarity. The problem was well recognized in early
studies\cite{Fried:90,Bagchi:91,DMcp:93} which suggested to use only
the longitudinal component of the field $\mathbf{\tilde
  E}_0(\mathbf{k})$. As a matter of fact, the problem lies in the
response function of the dipolar polarization field which needs to be
re-normalized with the account of the solute repulsive core, a
procedure similar to applying boundary conditions to the Poisson
equation of continuum electrostatics.

The Li-Kardar-Chandler\cite{Li:92,Chandler:93} Gaussian model allows
one to achieve a correct renormalization of the inhomogeneous
polarization response function $\bm{\chi}_p(\mathbf{k}_1,\mathbf{k}_2)$
eliminating the ``transverse catastrophe''.\cite{DMjcp1:04} This
approach introduces another simplification by replacing all the
short-range solute-solvent interactions by hard-core repulsions. This
simplification, however, leads to an exact solution for the
$\mathbf{k}$-space response functions with the result:\cite{DMjcp1:04}
\begin{equation}
  \label{eq:1-7}
  \bm{\chi}_p(\mathbf{k}_1,\mathbf{k}_2) = \bm{\chi}_s(\mathbf{k}_1)\delta_{\mathbf{k}_1,\mathbf{k}_2} -
     \bm{\chi}''(\mathbf{k}_1)\theta_0(\mathbf{k}_1-\mathbf{k}_2)\bm{\chi}_s(\mathbf{k}_2) .
\end{equation}
The second summand in Eq.\ (\ref{eq:1-7}) is the correction of the
response function of the homogeneous solvent, appearing in Eq.\
(\ref{eq:1-6-2}), by the repulsive core of the solute.  The response
function $\bm{\chi}''(\mathbf{k}_1)$ then incorporates both 
$\bm{\chi}_s$ and the information about the solute
shape.\cite{DMjcp1:04,DMjcp2:04}

A direct substitution of Eq.\ \eqref{eq:1-7} into Eq.\ \eqref{eq:1-3}
results in a 6D integral convolution in $\mathbf{k}$-space which is
not numerically tractable. In order to arrive at a computationally
efficient procedure, a mean-field approximation was
introduced,\cite{DMjcp2:04} which replaces the inhomogeneous electric
field of the solvent inside the solute by a mean cavity field:
\begin{equation}
  \label{eq:1-8}
  \mathbf{F}_0 = \frac{f}{8\pi} 
        \int_{\Omega} \mathbf{E}_0\cdot\mathbf{D}_{\mathbf{r}}\frac{d\mathbf{r}}{r^3} ,
\end{equation}
where
\begin{equation}
  \label{eq:1-9}
  f = \frac{2(\epsilon_s-1)}{2\epsilon_s+1} .
\end{equation}
Here, $\mathbf{D}_{\mathbf{r}}=3\mathbf{\hat r}\mathbf{\hat r} -
\mathbf{1}$ is the 2-rank dipolar tensor with $\mathbf{\hat
  r}=\mathbf{r}/r$.  $\mathbf{F}_0$ becomes the Onsager reaction
field\cite{Onsager:36} for a spherical solute with point dipole
located at the center.

The mean-field approximation reduces the problem of calculating the
solvation thermodynamics to a numerically tractable 3D integral in
$\mathbf{k}$-space. The chemical potential of solvation then becomes a
sum of two components arising from the longitudinal (L) and transverse
(T) polarization fluctuations, $\mu_{0s}^{L,T}$, and a third component
arising from the density fluctuations, $\mu_{0s}^d$:
\begin{equation}
  \label{eq:1-10}
  \mu_{0s} = \mu_{0s}^L + \mu_{0s}^T + \mu_{0s}^d .
\end{equation}

The transverse component $\mu_{0s}^T$ is defined by
$\mathbf{k}$-integral of the transverse projection of the solute
filed, $\tilde E_0^T(\mathbf{k})$, with the transverse polarization
structure factor
\begin{equation}
  \label{eq:1-11}
 - \mu_{0s}^{T} =  g_{\text{K}}^{-1}S^L(0)\frac{3y}{8\pi} \int \frac{d\mathbf{k}}{(2\pi)^3} 
                         |\tilde E^{T}_0(\mathbf{k})|^2 S^{T}(k) .
\end{equation}
The transverse field component is defined by subtracting the longitudinal
projection
\begin{equation}
  \label{eq:1-12}
  \mathbf{\tilde E}_0^L(\mathbf{k}) 
         = \mathbf{\hat k}\left(\mathbf{\hat k}\cdot\mathbf{\tilde E}_0\right)
\end{equation}
from the total inverted-space field $\mathbf{\tilde E}_0$:
\begin{equation}
  \label{eq:1-13}
  \mathbf{\tilde E}_0^T(\mathbf{k}) = \mathbf{\tilde E}_0(\mathbf{k})
   -  \mathbf{\tilde E}_0^L(\mathbf{k}) .
\end{equation}

Equation (\ref{eq:1-11}) is the main result of the application of the
Gaussian model\cite{Chandler:93} to polar solvation. It replaces
$S^T(k)$ of the direct perturbation expansion in Eqs.\
(\ref{eq:1-6-1}) and (\ref{eq:1-7}) with the renormalized function:
\begin{equation}
  \label{eq:1-13-1}
  3yS^T(k) \to (3yS^L(0)/g_K) S^T(k) .
\end{equation}
The $k=0$ limit of the transverse response changes from $3yS^T(0)=\epsilon
_s-1$ to $3(\epsilon_s-1)/(2\epsilon_s +1)$ thus eliminating the ``transverse
catastrophe'' of direct perturbation expansions.
 
The component $\mu_{0s}^L$ of the solvation chemical potential in Eq.\
(\ref{eq:1-10}) is obtained by inverted-space integration with the
longitudinal polarization structure factor:
\begin{equation}
  \label{eq:1-15}
   -\mu_{0s}^L = \frac{3y}{8\pi} \int \frac{d\mathbf{k}}{(2\pi)^3}S^L(k)
        \left[ |\tilde E_0^L|^2 - |\tilde E_0^T|^2 f 
   \frac{\mathbf{F}_0\cdot\mathbf{\tilde E}_0^L}{\mathbf{F}_0\cdot\mathbf{\tilde E}_0^T}\right] .
\end{equation}
There is a significant physics behind the appearance of the
transverse field in the brackets of Eq.\ \eqref{eq:1-15}. Longitudinal
dipolar polarization is short-ranged and thus does not propagate over
macroscopic distances. On the contrary, transverse polarization is
long-ranged. Therefore, inducing transverse polarization modifies the
electric field acting on the solvent dipoles resulting in the second
term in the brackets in Eq.\ \eqref{eq:1-15}.

\begin{figure}[htbp]
  \centering
  \includegraphics*[width=7cm]{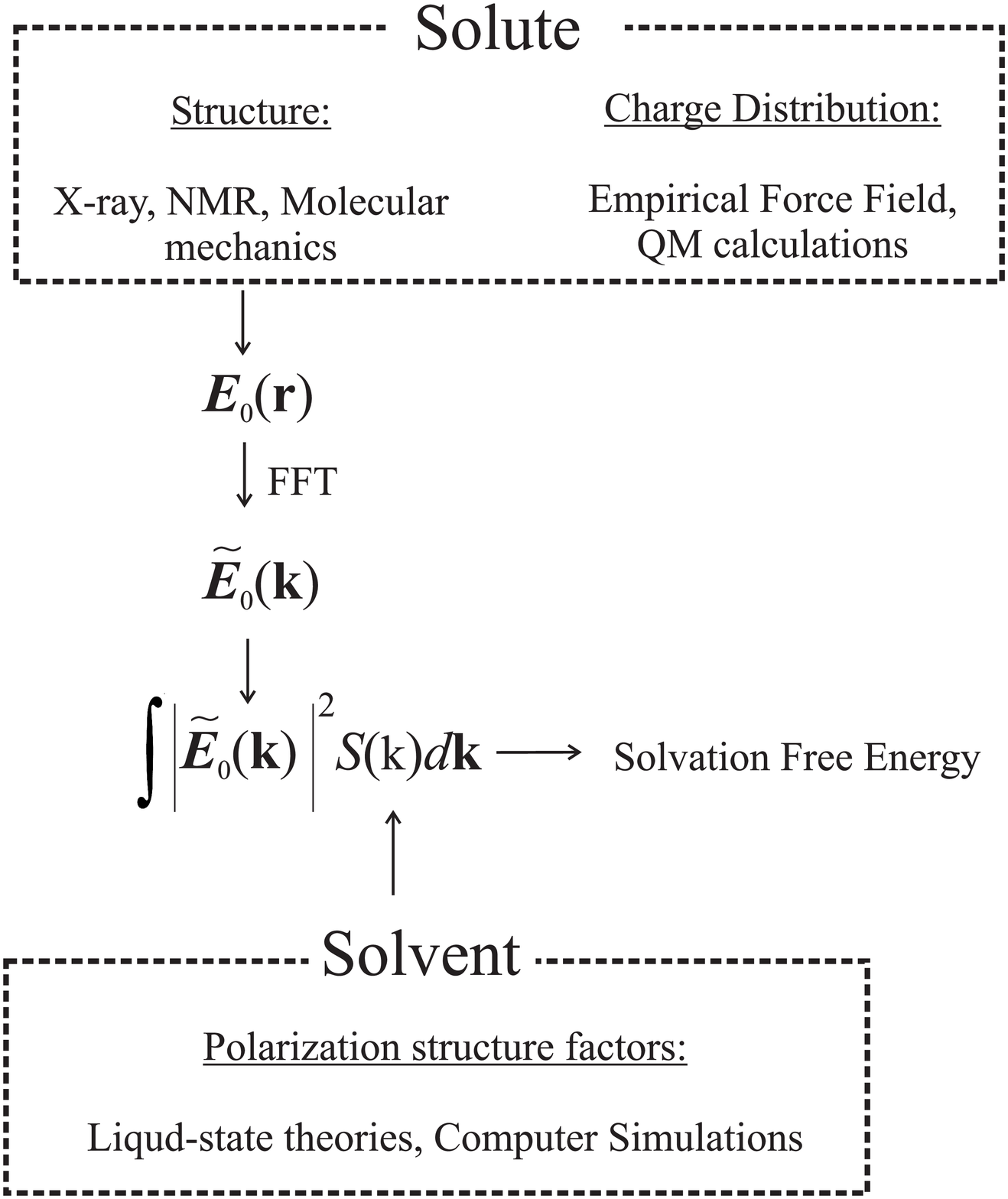}
  \caption{Diagram of the computational algorithm.}
  \label{fig:3}
\end{figure}

The density component in Eq.\ \eqref{eq:1-10} can formally be obtained
by multiplying the response function
$\bm{\chi}_d(\mathbf{k}_1,\mathbf{k}_2)$ [Eq.\ \eqref{eq:1-6-3}] with the
Fourier transforms of the electric field and integrating over
$\mathbf{k}_1$ and $\mathbf{k}_2$. This, however, results in a 6D
convolution integral to be avoided in numerical applications. An
alternative approach is to use direct-space integration when the density
component becomes
\begin{equation}
  \label{eq:1-15-1}
  -\mu_{0s}^d = 3 y F(\mathbf{r})*\mathcal{F}^{-1}\left[(1-S(k))\theta_0(\mathbf{k})\right] .
\end{equation}
Here, $F(\mathbf{r})=(8\pi)^{-1}E_0^2(\mathbf{r})$ is the density of the
electrostatic field energy and the asterisk indicates integration in
real space over the volume $\Omega$ occupied by the solvent. In addition,
$\mathcal{F}^{-1}$ is the inverse Fourier transform of the function
indicated in the brackets.

Solvation by the overall dipolar polarization, including nuclear and
electronic components, was considered in the formalism outlined above.
For solvation problems relevant to spectroscopy and charge-transfer
reactions nuclear component of polarization needs to be extracted.
This is achieved by replacing the density $y$ [Eq.\ (\ref{eq:1-6})] of
all, permanent and induced, dipoles in the equations above with the
density of permanent dipoles only, $y\to y_p=(4\pi/9)\beta m'^2 \rho$. In
addition, the $k=0$ values of the structure factors need to be
modified to account for screening of the dipolar interactions by the
high-frequency dielectric constant $\epsilon_{\infty}$. The $k=0$ values for these
nuclear structure factors, $S_n^{L,T}(k)$, now become\cite{DMcp:06}
\begin{equation}
  \label{eq:1-20}
  \begin{split}
    S_n^L(0) & = (\epsilon_{\infty}^{-1} - \epsilon_s^{-1}) /(3y_p), \\
    S_n^T(0) & = (\epsilon_s - \epsilon_{\infty}) /(3y_p).
  \end{split}
\end{equation}
Once the $k=0$ values for the structure factors are fixed by Eqs.\
(\ref{eq:1-14-1}) and (\ref{eq:1-20}), the scalar functions $S^L(k)$
and $S^T(k)$ can be calculated from our parametrization scheme,
parametrized polarization structure factors (PPSF).\cite{DMjcp2:04}
This analytical route to the polarization structure factors is tested
here by comparing the results of solvation calculations employing the
PPSF to the direct use of $S^{L,T}(k)$ from MD simulations (see
below).

\section{Computational algorithm}
\label{sec:3}
The computational algorithm is outlined in Fig.\ \ref{fig:3}.  The
solute is parametrized by coordinates $\mathbf{r}_j$, vdW radii $a_j$,
and partial charges $q_j$ of the atoms. The electric field of the
solute is calculated at points $\mathbf{r}_n$ of the $N\times N\times N$ grid
built on the $L\times L\times L$ cube:
\begin{equation}
  \label{eq:2-1}
  \mathbf{E}_0(\mathbf{r}_n) = \sum_{j=1}^{N_q} 
       \frac{q_j(\mathbf{r}_n-\mathbf{r}_j)}{|\mathbf{r}_n-\mathbf{r}_j|^3} ,
\end{equation}
where $N_q$ is the number of solute charges. The array
$\mathbf{E}_0(\mathbf{r}_n)$ is converted to inverted space by using
fast Fourier transform technique.\cite{Fortran:96} The field
$\mathbf{\tilde E}_0(\mathbf{k})$ is split into longitudinal and
transverse components and used in $\mathbf{k}$-integration in Eqs.\
\eqref{eq:1-11} and \eqref{eq:1-15} with the corresponding structure
factors of the dipolar polarization. As is illustrated in Fig.\
\ref{fig:3}, the calculation input is subdivided into two separate
components related to the solute and solvent properties. Details of
calculation for each of these are given below.

\subsection{Solute}
\label{sec:3-1}
\begin{figure}[htbp]
  \centering \includegraphics*[width=6cm]{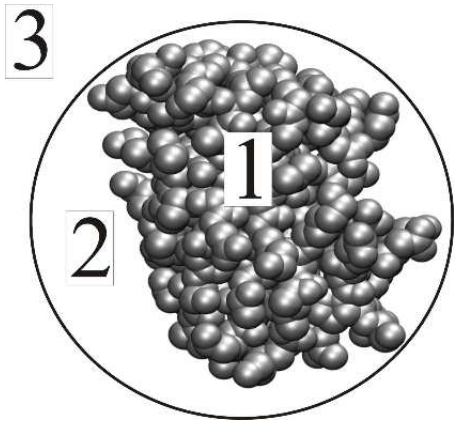}
  \caption{Separation of real space into regions for the
    calculation of the Fourier transform of the solute electric field
    [Eq.\ \eqref{eq:2-2}]. The Fourier transform is calculated numerically
    in region 2 and analytically [Eq.\ \eqref{eq:2-6}] in region 3.
    The field is set equal to zero within the hard repulsive core of the
    solute (region 1).}
  \label{fig:4}
\end{figure}

The Fourier transform of the Coulomb field is conditionally
convergent.  Therefore, in order to avoid numerical divergence in the
Fourier integral, real space is divided into three regions: hard core
of the solute (region 1), region outside a sphere of radius $R$
(region 3), and the region between the solute surface and the sphere
(region 2, Fig.\ \ref{fig:4}). The Fourier integral then becomes
\begin{equation}
  \label{eq:2-2}
  \mathbf{\tilde E}_0(\mathbf{k}) = \int_{r<R}\mathbf{E}_0(\mathbf{r})  
   e^{-i\mathbf{k}\cdot\mathbf{r}}d\mathbf{r} + \mathbf{\tilde E}_R,
\end{equation}
where the first integral is taken over region 2 and
\begin{equation}
  \label{eq:2-3}
  \mathbf{\tilde E}_R(\mathbf{k}) = 
          \int_{r>R}  \mathbf{E}_0(\mathbf{r})  e^{-i\mathbf{k}\cdot\mathbf{r}}d\mathbf{r} .
\end{equation}
The center of the sphere is taken at the
center of the charge distribution defined by the relation
\begin{equation}
  \label{eq:2-5}
  \mathbf{r}_q = \frac{\sum_{j=1}^{N_q}q_j\mathbf{r}_j}{\sum_{j=1}^{N_q} q_j}.
\end{equation}
The radius of the cut-off sphere $R$ is chosen to minimize the part of
the grid which is used in numerical calculation of the Fourier
transform.  In our calculations, the radius $R$ is chosen by adding
the solvent diameter $\sigma$ to the largest distance from $\mathbf{r}_q$
to the solvent-accessible surface (SAS) of the solute (vdW radii of
the surface atoms plus the radius of the solvent molecule).

The Fourier transform outside the sphere can be evaluated analytically.
For the location of charges relative to the center of charge given as
$\mathbf{s}_j = \mathbf{r}_j -\mathbf{r}_q$, the solution for
$\mathbf{\tilde E}_R$ can be obtained by expanding $\mathbf{E}_0(\mathbf{r})$
in $s_j/R<1$:
\begin{equation}
  \label{eq:2-6}
   \begin{split}
  \mathbf{\tilde E}_R(\mathbf{k}) =& -4\pi e\sum_j q_j \sum_{n=1}^{\infty} \left(\frac{s_j}{R}\right)^{n-1}
              \frac{j_{n-1}(kR)}{k}\\
& \left[ \mathbf{\hat s}_j P'_{n-1}(\cos\theta_j) - \mathbf{\hat k}P_{n}'(\cos \theta_j ) \right].
  \end{split}
\end{equation}
Here, $\cos \theta_j = \mathbf{\hat s}_j\cdot\mathbf{\hat k}$, $j_n(x)$ is the
spherical Bessel function, and $P_n(\cos \theta_j)$ is the Legendre
polynomial.

\begin{table*}
  \centering
  \caption{{\label{tab:1}}Atomic partial charges for copper and its four 
    ligands in the reduced (Red) and oxidized (Ox) states of PC.}
 \begin{ruledtabular}
  \begin{tabular}{ccccccccccc}
Set & \multicolumn{5}{c}{Red}  &  \multicolumn{5}{c}{Ox}  \\
    &  Cu  & N$_{\delta}$\footnotemark[1] & N$_{\delta}$\footnotemark[2] & S$_{\gamma}$\footnotemark[3] & 
             S$_{\delta}$\footnotemark[4] &
       Cu  & N$_{\delta}$\footnotemark[1] & N$_{\delta}$\footnotemark[2]  & S$_{\gamma}$\footnotemark[3] 
           & S$_{\delta}$\footnotemark[4]\\
\hline
I   &  1.0 & $-0.7$ &  $-0.7$ & $-1.23$ & $-0.09$ &
       2.0 & $-0.7$ &  $-0.7$ & $-1.23$ & $-0.09$   \\
II  &  $-0.49$ & $-0.445$ &  $-0.495$ & $-0.369$ & $-0.24$ &
       0.35 & $-0.42$ &  $-0.47$ & $-0.26$ & $-0.24$ \\
  \end{tabular}
 \end{ruledtabular}
\footnotetext[1]{His87}
\footnotetext[2]{His37}
\footnotetext[3]{Cys84 }
\footnotetext[4]{Met92}
\end{table*}

\subsection{Charging Scheme}
\label{sec:3-2}
The formal charge of the copper ion is +2 and of the cysteine sulfur
is $-1$ in the oxidized state of PC. The charge is, however,
delocalized among the ligands and the metal center. The main factor in
this delocalization is a strong covalency of a copper-sulfur (Cys)
bond. Calculations by Solomon and co-workers\cite{Solomon:93} assign
40\% of spin density of an unpaired electron to copper and 36\% to
cystein's sulfur in the ground state of oxidized protein. The extent
of delocalization varies significantly depending on the level of
quantum mechanical calculations
used.\cite{Ungar:97,Ullmann:97,Comba:02} The electron-nuclear
double resonance (ENDOR) experiments,\cite{Werst:91} which require
additional calibration on quantum calculations, result in the
following net charges on the residues coordinating
copper:\cite{Libeu:97} $-0.25$ (His), $-0.51$ (Cys), and $-0.04$
(Met). The more recent mapping of the electron spin density to NMR
relaxation\cite{Hansen:04} gives a much lower extent of
delocalization: $-0.11$ (Cys), $-0.025$ (His), 0 (Met).

The uncertainties in the extent of electron delocalization pose the
question of their impact on the calculation of the redox thermodynamics.
In order to study this question, we have performed calculations of the
solvation part of the redox potential and the corresponding entropy
using different charge sets. Set I is chemically fake assuming charge
$+2$ on copper in Ox state and the net charge of $-1$ on cysteine.
The negative charge is placed on cysteines sulfur in addition to
$-0.23$ from CHARMm22 protein parametrization. The rest of the
protein charges are from the standard CHARMm parametrization. The
reduced state for Set I is obtained by changing the metal charge to
$+1$. The charges for copper and its four ligands are summarized in
Table \ref{tab:1}. Set II is based on the charging scheme listed by
Ullmann \textit{et al}.\cite{Ullmann:97} for the oxidized state of
PC. The reduced form is obtained by placing an extra negative charge
on copper and its three ligands, $N_{\delta}$ (His87), $N_{\delta}$ (His37),
$S_{\gamma}$(Cys84), and $S_{\gamma}$(Met92) in proportion extracted from NMR
experiments (Table \ref{tab:1}).\cite{Hansen:04} Finally, a third
charge distribution is completely parametrized at the DFT level for
the charges and force constants of the copper and ligand atoms and
consistent with the Amber force field.\cite{realq299} In addition,
Amber FF03 parametrization \cite{amberFF03} was applied to all
non-ligand residues (Set II).  There were various numbers of TIP3P
water molecules for each of the charge distributions: 5,874 (Set I),
5,886 (Set II), and 4,628 (Set III).

We ran separate simulations (ca.\ 5 ns) for each charging scheme to
find that the results are not strongly affected by the choice of
atomic charges (Table \ref{tab:2}).  This was also noticed in some
other recent simulations.\cite{Cascella:06,Blumberger:06} We have
therefore implemented charge scheme II in all simulations reported
here since it presents a reasonable balance between being simple and
realistic.

\begin{table*}

  \caption{Temperature dependent $\langle V_{0s}\rangle_{\text{Ox}} $ (eV) for PC(Ox). The results are
    obtained from MD simulations, DelPhi calculations (with vdW and SS
  cavities), and from NRFT calculations.  The calculations were done
  with three charge distributions
  of the active site (I-III, see text for description).}
  \label{tab:2}
  \begin{ruledtabular}
  \begin{tabular}{ccccccccccc}
& \multicolumn{3}{c}{MD}  & \multicolumn{3}{c}{DelPhi(vdW)}  & \multicolumn{3}{c}{DelPhi(SS)}  & NRFT\footnotemark[1]  \\
\hline
   T/K  &  I  & II & III & I  & II & III & I  & II & III & II \\
\hline
  $280$ & $-68.29$ & $-69.46$ & $-68.37$ & $-104.86$ & $-102.07$ &
  $-99.76$ & $-46.90$ & $-47.47$ & $-46.37$ & $-104.6(-37.2)$  \\
  $285$ & & $-69.69$ &  & $-104.81$ & $-102.03$ & $-99.71$ & $-46.88$
  & $-47.45$ & $-46.36$ 
& $-103.3(-36.5)$ \\
  $290$ & $-70.73$ & $-66.01$   &  & $-104.76$ & $-101.98$  & $-99.67$
  & $-46.87$ & $-47.43$ & $-46.34$ 
           & $-102.1(-35.8)$ \\
  $295$ & & $-66.97$   &  & $-104.71$ & $-101.93$ & $-99.62$ &
  $-46.84$ & $-47.41$ & $-46.32$ 
           & $-100.7(-35.1)$ \\
  $300$ & $-67.06$ & $-66.09$   & $-65.84$ & $-104.65$ & $-101.88$ & $-99.57$ & $-46.81$ & $-47.39$ & $-46.30$ 
           & $-99.3(-34.5)$\\
  $305$ & & $-68.58$   &  & $-104.59$ & $-101.82$ & $-99.52$ &
  $-46.80$ & $-47.37$ & $-46.28$ 
            & $-98.3(-33.9)$ \\
  $310$ & & $-67.51$  & $-66.68$ & $-104.53$ & $-101.77$ & $-99.46$ &
  $-46.79$ & $-47.35$ & $-46.26$ 
            & $-97.1(-33.3)$  \\
\end{tabular}
\end{ruledtabular}
\footnotetext[1]{The numbers in the parentheses indicate the density
  component of the equilibrium interaction energy, $\langle V_{0s}\rangle_{\text{Ox}}^d$. }
\end{table*}

\subsection{Solvent}
The polarization structure factors entering the equations for the
solvation chemical potential are characteristics of the homogeneous
solvent.  They can be obtained numerically by averaging the
projections of dipole moments $\mathbf{\hat e}_j$ on an arbitrary
chosen direction of the $\mathbf{k}$-vector, $\mathbf{\hat k}=\mathbf{k}/k$:
\begin{equation}
  \label{eq:2-7}
  \begin{split}
    S^L(k) &= \frac{3}{N}\left\langle \sum_{i,j} (\mathbf{\hat e}_j \cdot \mathbf{\hat k})
           (\mathbf{\hat k}\cdot \mathbf{\hat e}_i ) e^{i\mathbf{k}\cdot\mathbf{r}_{ij}} \right \rangle, \\
    S^T(k) &= \frac{3}{2N}\left \langle \sum_{i,j}\left[ (\mathbf{\hat e}_j\cdot \mathbf{\hat e}_i) - 
              (\mathbf{\hat e}_j \cdot \mathbf{\hat k}) 
             (\mathbf{\hat k}\cdot \mathbf{\hat e}_i ) \right]  
            e^{i\mathbf{k}\cdot\mathbf{r}_{ij}} \right\rangle, 
  \end{split}
\end{equation}
where $\mathbf{r}_{ij}=\mathbf{r}_i - \mathbf{r}_j$ and $N$ is the
number of liquid dipoles.

\begin{figure}
  \centering
  \includegraphics*[width=6.5cm]{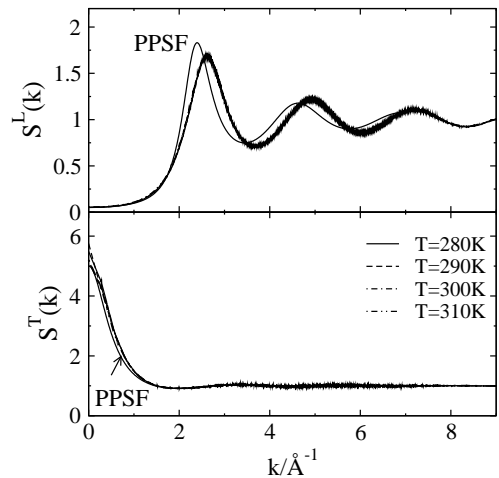}  
  \caption{Longitudinal (L) and transverse (T) polarization 
           structure factors of TIP3P water [Eq.\ \eqref{eq:2-7}] calculated
           at different temperatures from $\simeq 25$ ns MD
           trajectories. Also shown is the PPSF calculation at $T=300$ K.
           }
  \label{fig:5}
\end{figure}

Unfortunately, experiment does not provide spatially resolved
correlators of dipoles in polar liquids and one has to resort to using
either computer simulations or liquid-state theories.  Parametrizing
homogeneous structure factors by computer simulations is a very
attractive avenue for studying hydration because of continuously
improving empirical potentials for water\cite{Guillot:02} on one hand
and the reliance of biological applications on aqueous solvation on
the other.  For the problems related to derivatives of thermodynamic
potentials, e.\ g.\ entropy and volume of solvation, the structure
factors need to be tabulated at different temperatures and/or
pressures.  In this paper, polarization structure factors $S^{L,T}(k)$
were obtained from $\simeq 25$ ns MD trajectories of TIP3P
water\cite{tip3p:83} at different temperatures (Fig.\ \ref{fig:5}).

In parallel to simulations, we have used dipolar structure factors
from our PPSF parametrization scheme.\cite{DMjcp2:04} This approach is
based on the analytical solution of the mean-spherical approximation
for the fluid of dipolar hard spheres\cite{Wertheim:71} which is
parametrized to give $k=0$ values from Eqs.\ \eqref{eq:1-14-1} and
(\ref{eq:1-20}).\cite{DMjcp2:04,DMcp:06} The PPSF parametrization
gives solvation free energies essentially identical to those obtained
with the structure factors from simulations, as was also found for a
smaller polypeptide solute in our previous publication.\cite{DMcp:06}
It appears therefore that the local tetrahedral order of water, which
is of course not captured by dipolar hard spheres, is not significant
for the energetics of polar solvation dominated by orientational
correlations ruled by dipole-dipole forces.  The density fluctuations
are, on the other hand, dominated by repulsions. We used, therefore,
the density structure factor from the Percus-Yevick
solution\cite{Hansen:03} for a hard-sphere fluid of the same density
as water in calculations of the density response in Eq.\
(\ref{eq:1-15-1}).

We need to emphasize here that using dielectric constants of model
dipolar fluids would give us wrong results. The dielectric constant of
a molecular liquid is affected by short range molecular correlations
through the Kirkwood factor, hydrogen bonds and molecular quadrupoles
are among significant factors.\cite{SPH:81} We account for all these
effects in the PPSF scheme by using the experimental, either from
computer or laboratory data, dielectric constants in Eqs.\
(\ref{eq:1-14-1}) and (\ref{eq:1-20}). Once the $k=0$ limit is set up
by the experimental input, the behavior of $S^{L,T}(k)$ in the range
of $k < 2\pi / \sigma $ ($\sigma$ is the solvent diameter) is well reproduced by
solutions obtained for model dipolar fluids.  These solutions will
fail at $k\geq 2\pi/ a$, where $a$ is the characteristic distance between
partial charges within the solvent molecule.  This part of the
spectrum of polarization fluctuations is correctly captured by models
based on interaction-site integral equations,\cite{Raineri:99} but
that range of wave-vectors normally does not contribute to the
solvation energy. In fact, the range of $k$-values relevant for the
solvation problem is limited by $k<2\pi / R$, where $R$ is the
characteristic dimension of the solute. For large solutes, only the
long-wavelength part of the polarization structure factors is really
needed for the solvation energy calculations. As is shown in Fig.\
\ref{fig:5}, there is a mismatch between the PPSF longitudinal
structure factor and MD simulations. However, this difference makes no
effect on the calculated solvation energies.

We can summarize our results on parametrizing the solvent properties
by stating that the model fluid of dipolar hard spheres can serve as a
reliable reference system for calculations of polar solvation given
the macroscopic properties, the density of dipoles $y$ and the
dielectric constant $\epsilon_s$, have been taken from experiment
(either laboratory or computer). The theory thus adds an additional
parameter $y$ to the dielectric constant used in electrostatic
solvation theories to produce a fully microscopic solvent response.
In practical applications of the theory (e.g.\ in case of solvation in
ambient water presented below), the parameter $y$ needs to be
calculated from the molecular properties of the solvent. We use the
1-R Wertheim theory\cite{Wertheim:79} to calculate the effective
dipole moment of the solvent (see Ref.\ \onlinecite{DMjpca:04} for
comparison to simulations). The solvent input is thus made by five
parameters: $\{\sigma, \rho, m, \alpha, \epsilon_s\}$. One needs in addition the
high-frequency dielectric constant $\epsilon_{\infty}$ for the
reorganization energy calculations and the temperature slopes of two
dielectric constants, as well as the isobaric expansivity, for the
solvation entropy calculations. The big advantage of the PPSF scheme
is that all these parameters have been tabulated for many solvents
commonly used in solution chemistry making our method broadly
applicable to solvation calculations in polar molecular solvents.
 
Despite the fact that the dielectric constant is sensitive to local
correlations, the polarization structure factors in the
long-wavelength limit are fully determined by dipolar correlations
general for all polar liquids and not much sensitive to details of the
local structure which is of course very different in water than in a
hard-sphere dipolar fluid.  There are several advantages to using
dipolar hard spheres as the reference system. First, all thermodynamic
and structural properties are controlled by only two parameters, the
reduced density $\rho \sigma^3$ and the dipolar density $y$. Second, this
system is well characterized both analytically and numerically. It has
served many times as a starting point for developing theories of polar
liquids,\cite{Hansen:03} similarly to the role played by the fluid of
hard spheres in theories of non-polar liquids.\cite{WCA} Once that
stated, we however want to stress that the theory itself is based on
the structure factors of an arbitrary polar medium with the Gaussian
fluctuation spectrum and is not limited to a choice of any particular
reference system.

\section{Simulations protocol}
\label{sec:4}
Amber 8.0\cite{amber8} was used for all MD simulations.  The initial
configuration of PC was created using a protonated version of the
X-ray crystal structure at 1.7 \AA{} resolution (PDB: 1ag6\cite{1ag698}).
This initial configuration of the protein was first minimized in
vacuum by the conjugate gradient method for 10,000 steps to allow the
protein to remove any bad initial contacts.  Then the system was
solvated in a rectilinear box with several thousand TIP3P
molecules,\cite{tip3p:83} providing at least two-three solvation
shells around the protein.  To neutralize the charge, a number of
sodium ions equal to the total charge of the protein were added. The
protein was then relaxed for a few thousand steps while water and
sodium were positionally constrained.  Finally, the entire system
containing solvent, counterions, and protein was energy minimized in
100,000 steps.
  
Next, the system was heated in a NVT ensemble for 30 ps from 0 K to
the desired temperature followed by volume expansion in a 1 ns NPT
run. NVT production runs, following density equilibration, lasted from
6 to 18 ns. The last 5--10 ns at the end of each trajectory were used
to calculate the averages.  The timestep for all MD simulations was 2
fs, and SHAKE was employed to constrain bonds to hydrogen atoms.
Constant pressure and temperature simulations employed Berendsen
barostat and thermostat, respectively. \cite{berend84} The long-range
electrostatic interactions were handled using a smooth particle mesh
Ewald summation with a $9$ \AA{} limit in the direct space sum.  The total
charge for the protein was $-9.0$ for the reduced state and $-8.0$ for
the oxidized state.

\section{Results}
\label{sec:5}
The calculations presented here are focused on two properties: the
solvent portion of the redox chemical potential, $\Delta \mu_{s}$, and the
solvent reorganization energy $\lambda_s$, both corresponding to the half
reaction
\begin{equation}
  \label{eq:1-16}
  \mathrm{PC(Ox)}^{8-} + \mathrm{e}^{-} \to \mathrm{PC(Red)}^{9-} .
\end{equation}
The former can in principle be calculated as the difference of
solvation chemical potentials in the Red and Ox states. However, this
approach involves calculating the difference in two large numbers,
which is computationally unreliable. Instead, we use the linear
response approximation to calculate $\Delta \mu_s$ according to the equation:
\begin{equation}
  \label{eq:1-17}
  \Delta \mu_{s} =\mu_{0s}^{\text{Red}} - \mu_{0s}^{\text{Ox}} = - \Delta
  \mathbf{\tilde E}_0 * \bm{\chi} * \mathbf{\bar E}_0  .
\end{equation}
Here, $\mathbf{\bar E}_0 =
(\mathbf{\tilde E}_0^{\text{Ox}}+\mathbf{\tilde E}_0^{\text{Red}})/2$
and $\Delta\mathbf{\tilde E}_0 =
\mathbf{\tilde E}_0^{\text{Red}}-\mathbf{\tilde E}_0^{\text{Ox}}$ are the mean and the
difference of the electric fields in the Red and Ox states.  Similarly,
the solvent reorganization energy is calculated from
\begin{equation}
  \label{eq:1-18}
  \lambda_s = \frac{1}{2} \Delta \mathbf{\tilde E}_0 * \bm{\chi} * \Delta \mathbf{\tilde E}_0 .
\end{equation}
Equation \eqref{eq:1-18} applies to the reorganization energy of
non-polarizable solvents employed in computer simulations. For
laboratory data, nuclear polarization should be separated from the
overall solvent polarization and the response function $\bm{\chi}$ is
replaced by the nuclear response function $\bm{\chi}_n$ as explained
above and in more detail in Ref.\ \onlinecite{DMcp:06}.

\subsection{Redox Thermodynamics}
\label{sec:5-1}
The solvation thermodynamics calculated here can be related to
experimental redox entropies reported by measuring the temperature
dependence of the standard or midpoint electrode
potentials.\cite{Yee:79,Sailasuta:79,Sola:99} An electrochemical
experiment corresponds to bringing a solution containing given numbers
of oxidized and reduced reagents, which are not necessarily in
equilibrium (the ratio of their numbers is not a Boltzmann factor), in
contact with a metal electrode. The equilibrium is established between
the electronic subsystem of the redox pair and the electrode in such a
way that the electrode is charged and its electrochemical potential
$\mu$ is shifted from the vacuum Fermi energy $\epsilon_F$ by the electrostatic
potential $\phi$: $\mu = \epsilon_F - e\phi$. 

The numbers of the oxidized and reduced forms of the redox pair,
$N_{\text{Ox}}$ and $N_{\text{Red}}$, are assumed to be large enough
so that they are not affected by charging the electrode.  The
electrochemical potential of the electrode than becomes equal to the
absolute electrochemical potential of the redox couple in the
solution.\cite{Reiss:88} The latter can be found from simple
statistical arguments. The grand-canonical free energy of two
fermionic subsystems of $N_{\text{Ox}}$ and $N_{\text{Red}}$
electronic levels is\cite{Landau5}
\begin{equation}
  \label{eq:5-2}
 \begin{split}
  \beta \Omega & = -N_{\text{Ox}}\ln\left(1 + e^{\beta (\mu-\epsilon_{\text{Ox}})}\right)  \\
      & - N_{\text{Red}}\ln\left(1 + e^{\beta (\mu-\epsilon_{\text{Red}})} \right),
 \end{split}
\end{equation}
where $\epsilon_{\text{Ox}}$ and $\epsilon_{\text{Red}}$ are the average energies of
the electronic levels in the corresponding redox states. The chemical
potential is then found by requiring that the derivative $-(\partial \Omega/
\partial\mu)_T$ is equal to the total number of electrons $N_{\text{Red}}$. For
the energy gap between Ox and Red states greater than $k_{\text{B}}T$,
this requirement results in the Nernst equation\cite{Bockris:70}
\begin{equation}
  \label{eq:5-3}
  \mu = \frac{\epsilon_{\text{Ox}} + \epsilon_{\text{Red}}}{2} - k_{\text{B}}T \ln \left(N_{\text{Ox}}/N_{\text{Red}}\right), 
\end{equation}
in which the standard potential is given by the mean of the average 
electronic energies 
\begin{equation}
  \label{eq:5-4}
  \phi^0 = -\frac{\epsilon_{\text{Ox}} + \epsilon_{\text{Red}}}{2e} .
\end{equation}

\begin{table*}[htbp]

  \caption{Redox thermodynamics  of PC (eV). }
  \label{tab:3}
 \begin{ruledtabular}
  \begin{tabular}{cccccccccccc}
 & \multicolumn{4}{c}{$\Delta\mu_s$(TIP3P)}  &  \multicolumn{4}{c}{$\lambda_{s}$(TIP3P)} & 
   \multicolumn{3}{c}{$\lambda_{s}$(H$_2$O)}  \\
\hline
T/K  &  MD  & NRFT & vdW\footnotemark[1] & SAS\footnotemark[2] & 
        MD  & NRFT & vdW\footnotemark[3] &
        SAS & NRFT\footnotemark[4] &  vdW & SAS \\
\hline
$280$ & $-3.05$ & $-6.21$ & $-9.63$ & $-7.15$ & $0.845$ & $1.37(0.59)$ &
$3.645$ & $0.690$ & $0.82$ & $2.728$ & $0.417$ \\
$285$ & $-2.73$ & $-6.15$ & $-9.62$ & $-7.14$ & $0.814$ & $1.35(0.58)$ &
$3.641$ & $0.685$ & $0.80$ & $2.725$ & $0.417$ \\
$290$ & $-2.71$ & $-6.10$ & $-9.62$ & $-7.14$ & $0.646$ & $1.33(0.57)$ &
$3.636$ & $0.685$ & $0.79$ & $2.722$ & $0.417$ \\
$295$ & $-3.14$ & $-6.04$ & $-9.61$ & $-7.14$ & $0.565$ & $1.32(0.56)$ &
$3.630$ & $0.685$ & $0.78$ & $2.718$ & $0.417$ \\
$300$ & $-2.82$ & $-6.01$ & $-9.60$ & $-7.14$ & $0.435$ & $1.30(0.55)$
  & $3.623$ & $0.684$ & $0.77$ & $2.714$ & $0.417$ \\
$305$ & $-2.94$ & $-5.94$ & $-9.59$ & $-7.13$ & $0.426$ & $1.29(0.54)$ &
$3.620$ & $0.684$ & $0.76$ & $2.710$ & $0.417$ \\
$310$ & $-2.53$ & $-5.89$ & $-9.59$ & $-7.13$ & $0.543$ & $1.27(0.53)$ &
$3.618$ & $0.684$ & $0.74$ & $2.710$ & $0.416$ \\
\end{tabular}
\end{ruledtabular}
\footnotetext[1]{Poisson-Boltzmann calculations with vdW radii
  assigned to the protein atoms (standard vdW cavity).}
\footnotetext[2]{Poisson-Boltzmann calculation with solvent radius
  added to the radii of the protein atoms exposed to the solvent 
  (solvent-accessible cavity, SAS).}
\footnotetext[3]{The continuum calculations are done with the
  dielectric constant of TIP3P water $\epsilon_s=95$ and
  $\epsilon_{\infty}=1.0$. The dielectric constant of the protein interior was put
  equal to unity in order to be consistent with the microscopic
  calculations. The temperature variation of the dielectric constant
  of $d\epsilon_s/ dT = -0.654$ K$^{-1}$ (Ref.\ \onlinecite{DMcp:06}) was adopted for the entropy calculations
  listed in Tab.\ \ref{tab:4}. }
\footnotetext[4]{Calculations in ambient water using $\epsilon_{\infty}=1.78$, $\epsilon_s=78$, 
  $d\epsilon_s/dT=-0.398$ K$^{-1}$, and $d\epsilon_{\infty}/dT=-2.75\times 10^{-4}$ K$^{-1}$. In addition, 
  temperature expansion was included with the constant-pressure expansivity coefficient 
  $\alpha_p=2.6\times 10^{-4}$ K$^{-1}$.}
\end{table*}

\begin{figure}
  \centering
  \includegraphics*[width=6.5cm]{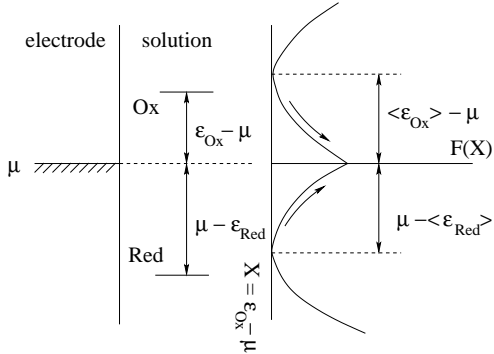}
  \caption{Contact of a redox pair with the metal electrode.
    $\epsilon_{\text{Ox}} -\mu $ and $\epsilon_{\text{Red}}-\mu $ show the fluctuating
    energy gaps for reduction and oxidation electron transfer,
    respectively. The equilibrium electrochemical potential of the
    electrode is established when the equilibrium energy gaps are
    equal for the reduction and oxidation reactions [Eq.\
    \eqref{eq:5-3}].  The Marcus electron transfer parabolas, shown by
    the dependence of free energy $G(X)$ on the energy gap coordinate
    $X=\epsilon_{\text{Ox}}-\mu$, are symmetric in this case producing equal
    oxidation and reduction currents [Eq.\ \eqref{eq:5-5}]. }
  \label{fig:6}
\end{figure}

The same result follows from the use of the stationary condition (zero
electrode current) for the rates of reduction and
oxidation\cite{Schmickler:96}
\begin{equation}
  \label{eq:5-5}
  k_{\text{Ox}} c_{\text{Ox}} = k_{\text{Red}} c_{\text{Red}} ,
\end{equation}
where $c_{\text{Ox/Red}}$ are the surface concentrations. By using the
Marcus equation for the reaction rate\cite{Marcus:65}
\begin{equation}
  \label{eq:5-6}
  k_{\text{Ox/Red}} \propto \exp\left[-\beta \frac{(\epsilon_{\text{Ox/Red}} - \mu)^2 }{4\lambda_s} \right]
\end{equation}
and neglecting the logarithmic correction including the ratio of two
surface concentrations, one gets equal rates at $ \mu \simeq (\epsilon_{\text{Ox}} +
\epsilon_{\text{Red}})/2$. The double-well Marcus free energy surface for the
electrode electron transfer is then symmetrical as illustrated in
Fig.\ \ref{fig:6}. This picture bears a clear similarity with the
formation of the Fermi level in the forbidden band of a semiconductor,
as was noticed by Reiss.\cite{Reiss:85}

The electronic energies are given by the sums of their vacuum
components, $\epsilon_{\text{Ox/Red}}^0$, and the interaction of the electric
field of the electron $\mathbf{E}_e$ with the polarization of the
solvent in equilibrium with the total electric field of the molecule
in the solution\cite{DMjec:91}
\begin{equation}
  \label{eq:5-7}
  \epsilon_{\text{Ox/Red}} = \epsilon^0_{\text{Ox/Red}} - \mathbf{E}_e*\bm{\chi}*\mathbf{E}_0^{\text{Ox/Red}} .
\end{equation}
Taking into account that 
\begin{equation}
  \label{eq:5-8}
  \mathbf{E}_e = \mathbf{E}_0^{\text{Red}} - \mathbf{E}_0^{\text{Ox}} = \Delta \mathbf{E}_0 ,
\end{equation}
one gets from Eqs.\ \eqref{eq:5-4}, \eqref{eq:5-7}, and \eqref{eq:5-8}
the commonly used connection between the standard electrode potential
and the solvation part of the redox free energy
\begin{equation}
  \label{eq:5-9}
  \phi^0 = - \frac{\epsilon^0_{\text{Red}} + \epsilon^0_{\text{Ox}}}{2e} - \frac{\Delta \mu_s}{e},
\end{equation}
where $\Delta \mu_s$ is given by Eq.\ \eqref{eq:1-17}. The first term in this
equation disappears in the temperature derivative reported
experimentally\cite{Sailasuta:79,Sola:99,Battistuzzi:03}
\begin{equation}
  \label{eq:5-11}
  e\left(\frac{\partial \phi^0}{\partial T}\right)_P =\Delta s_s = s^{\text{Red}}_s - s^{\text{Ox}}_s 
                                = - \left(\frac{\partial \Delta \mu_s}{\partial T}
                                \right)_P .
\end{equation}
We need to stress here that redox entropies in polar solutions are
sensitive to the presence of electrolyte.\cite{Swartz:96,Moore:86} One
therefore can expect only a qualitative agreement between experiments
done in buffered protein solutions\cite{Sola:99,Battistuzzi:03} and
our calculations at zero ionic strength.

From Eqs.\ \eqref{eq:5-4}, \eqref{eq:5-7}, and \eqref{eq:5-8} one can directly 
derive the equation for the solvation redox free energy
\begin{equation}
  \label{eq:5-12}
  \Delta \mu_s = \left(\langle\Delta V_{0s} \rangle_{\text{Ox}} + \langle\Delta V_{0s} \rangle_{\text{Red}}\right)/2 ,
\end{equation}
where $\Delta V_{0s}$ is the difference in the solute-solvent interaction
energies in the Red and Ox states and the averages are taken over the
corresponding ensembles.  The same average vertical gaps can be used
to calculate the reorganization energy as
\begin{equation}
  \label{eq:5-13}
  \lambda_s = \left(\langle\Delta V_{0s}\rangle_{\text{Ox}} - \langle\Delta V_{0s}
    \rangle_{\text{Red}}\right)/2 .
\end{equation}
We need to caution here is that while Eq.\ \eqref{eq:5-4} is a
statistical-mechanical result, Eqs.\ \eqref{eq:5-7}--\eqref{eq:5-13}
are based on the LRA for the solute-solvent interaction energy and
might be affected by deviations from this approximation.

\begin{figure}
  \centering
  \includegraphics*[width=6.5cm]{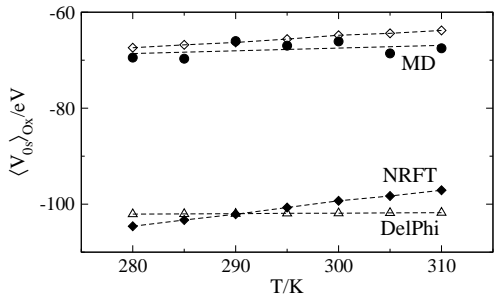}
  \caption{Average solute-solvent interaction energy $\langle V_{0s} \rangle_{\text{Ox}}$
    obtained from MD simulations (closed circles), NRFT (diamonds),
    and DelPhi continuum calculations (vdW cavity, triangles). The
    closed diamonds refer to the total average energy including the
    polarization and density components, while the open diamonds
    denote the polarization component only.  The dashed lines
    represent linear regressions through the points. }
  \label{fig:7}
\end{figure}

\subsection{Solute-solvent average energy}
In addition to our NRFT formalism, we have used the dielectric
continuum approximation implemented in the DelPhi program
suite\cite{delphi02} in the solvation calculations. Dielectric
constant of ambient water was used for the solvent continuum and $\epsilon_s
=1$ for the protein. This latter choice was driven by our desire to
compare continuum and microscopic calculations of solvation
thermodynamics since the latter does not assume any polarization of
the protein.  Table \ref{tab:2} lists the results of NRFT and DelPhi
calculations of the average energy $\langle V_{0s}\rangle_{\text{Ox}}$ of PC in
the Ox state. Three different charging schemes have been used and
compared to MD simulations (Table \ref{tab:1} and Sec.\
\ref{sec:4}). In the following we will discuss the results relevant to
charging scheme II only, which are also visualized in Fig.\
\ref{fig:7}.

\begin{figure}
  \centering
  \includegraphics*[width=6.5cm]{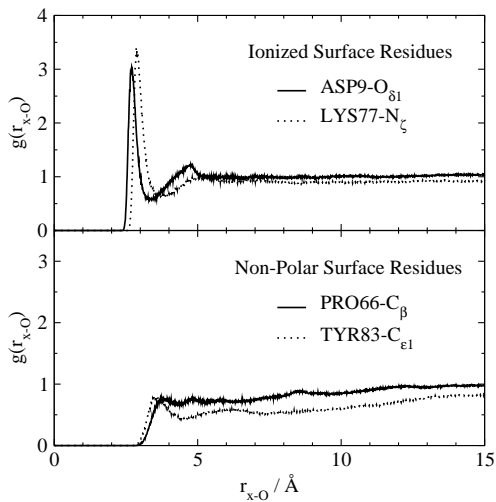}
  \caption{Radial distribution functions between surface residues of PC
    and oxygens of water. The upper panel shows ionized residues and
    the lower panel refers to non-polar residues. The legends in the
    figure list: aspartic acid (ASP), the probe atom is the oxygen at
    the 1st $\delta$ position; lysine (LYS) , the probe atom is nitrogen at
    the $\zeta$ position; proline (PRO), the probe atom is the $\beta$ carbon;
    tyrosine (TYR), with the first $\epsilon$ carbon as the probe atom. }
  \label{fig:8}
\end{figure}

The NRFT calculations listed in Table \ref{tab:2} and shown in Fig.\
\ref{fig:7} have been done by using Eqs.\ (\ref{eq:1-3}) and
(\ref{eq:1-6-1}) in which the electric field of PC in Ox state was
used for $\mathbf{\tilde E}_0(\mathbf{k})$. The close diamonds in
Fig.\ \ref{fig:7} refer to the total solvent response, while open
diamonds represent the polarization response only [$\bm{\chi}_p$ in Eq.\
(\ref{eq:1-6-1})]. Two interesting observations result from examining
Fig.\ \ref{fig:7}: (i) a close proximity of the NRFT result to the
standard (vdW) continuum calculation and (ii) a good agreement between
the polarization portion of the NRFT calculations and MD results. The
continuum electrostatics does not reproduce the slope of the average
energy as we also discuss below in relation to the redox entropy.

In order to understand the origin of the close agreement between MD
and the polarization component of the solvent response, one needs to
recall what comes to the calculation of the polarization and density
components of the solvation free energy. The polarization response is
calculated by assuming that the only influence of the solute on the
polarization field is to exclude it from the solute volume represented
by the step function $\theta_0(\mathbf{r})$ equal to one inside the solute
and zero otherwise [Eq.\ (\ref{eq:1-7})]. The density component
corrects this result by taking into account the inhomogeneous density
profile formed at the surface of a hard-wall solute. In dense liquids,
such a profile is characterized by a sharp peak of the radial
distribution function in the first solvation shell of the
solute. Correspondingly, reflecting the belief that the short-range
structure of liquids is primarily determined by repulsions,\cite{WCA}
the density structure factor $S(k)$ in Eq.\ (\ref{eq:1-15-1}) was
taken in our calculations from the Percus-Yevick solution for hard
spheres.\cite{Hansen:03} The close proximity of the full NRFT
calculation to the standard DelPhi/vdW algorithm (Fig.\ \ref{fig:7})
illustrates the fact that the common parametrization of the atomic
radii is based on the experience learned for hydration of small ions
with tightly bound first solvation shell. In the present algorithm,
this physics is accommodated by the density component of the solvation
free energy.

The structure of water at the protein surface is quite different from
what is normally obtained by inserting a small solute in a molecular
solvent. The structure is heterogeneous including islands of highly
structured water around polar and ionized residues combined with much
softer density profile at the hydrophobic patches.  This reality is
illustrated in Fig.\ \ref{fig:8} which shows pair distribution
functions between ionized and non-polar residues and water's oxygens.
While the distribution functions of ionized residues are reminiscent
of the structures typically observed around small solutes in dense
solvents, the water structure around non-polar residues is quite
different: there is no first-shell peak and water interface is shifted
by $\simeq 1$ \AA, in accord with simulations of nano-scale hydrophobic
solutes.\cite{Hotta:05}

The stronger attraction of the surface water molecules to the bulk
than to a non-polar hydrophobic patch of the protein (cavity expulsion
potential\cite{Stillinger:73,HummerPRL:98}) results in a weak
dewetting of the surface\cite{Choudhury:07} with the density at the
interface lower than in the bulk (Fig.\ \ref{fig:8}). Since there are
only a few charged residues on the protein surface, the average
surface structure is closer to a step-wise cut-off introduced in the
polarization component of the response function than to a structured
liquid at the surface of a small polar/ionic
solute.\cite{ChandlerNature:05} This observation explains a good
agreement between MD and polarization calculations of the solvation
thermodynamics in this paper as well as an equally impressive
agreement with the simulations obtained in our previous calculations
of charge transfer across a polypeptide bridge.\cite{DMcp:06}

\subsection{Solvent Gibbs and reorganization free energies}
\begin{figure}
  \centering
  \includegraphics*[width=6.5cm]{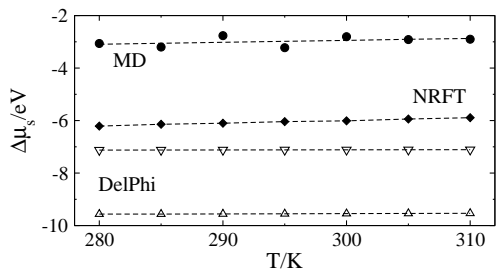}
  \caption{Solvation Gibbs energy from MD simulations (closed circles), NRFT
    calculations (closed diamonds), continuum DelPhi calculation with the
    standard cavity definition (up-triangles), and the cavity surface
    augmented by the solvent radius $\sigma/2$ (down triangles). The dashed
    lines are linear regressions through the points. }
  \label{fig:9}
\end{figure}

\begin{figure}
  \centering
  \includegraphics*[width=6.5cm]{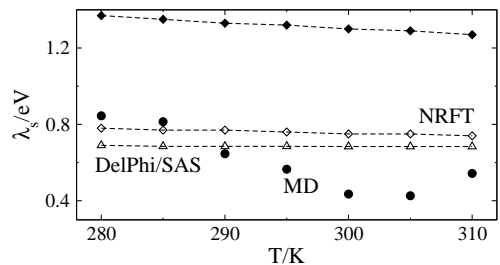}
  \caption{Reorganization energy of PC vs temperature calculated from
    MD simulations (closed circles), from NRFT (diamonds), and from
    dielectric continuum using solvent-accessible cavity definition
    (triangles). The dashed lines are linear regressions through the
    points.  The filled diamonds refer to the full NRFT calculation
    and the open diamonds denote the polarization response only.  }
  \label{fig:10}
\end{figure}

The results of calculations of the redox solvation energy and solvent
reorganization energy [Eqs.\ \eqref{eq:5-12} and \eqref{eq:5-13}]
using different levels of the theory are listed in Table \ref{tab:3}.
Redox and reorganization entropies are given in Table \ref{tab:4}. In
addition, the temperature dependence of $\Delta \mu_s$ and $\lambda_s$ are
visualized in Figs.\ \ref{fig:9} and \ref{fig:10}. The differences in
the theoretical results arise from the different level of structural
solvent information incorporated in each formalism. For completeness,
we have also listed in Table \ref{tab:3} the NRFT calculations with
the solvent parameters of ambient water.

The continuum electrostatics gain access to the solvation entropy
through the temperature variation of the solvent dielectric constant.
Therefore, since the dielectric constant commonly decreases with
heating, the solvent becomes effectively less polar and the solvation
free energy of a charge distribution increases, i.e.\ becomes less
negative. If one considers redox species positively charged in both
redox states, the oxidized state carries a larger charge and hence the
difference of Red and Ox solvation energies has a positive value
decreasing with increasing temperature.  The redox entropy in Eq.\
\eqref{eq:5-11} is then positive as is typically observed for simple
inorganic ions.\cite{Yee:79,Florian:99} By the same arguments, the
species carrying negative charge in both redox states should have a
negative redox entropy, which is the case for the negatively charged
PC in our calculations.

\begin{table*}[htbp]
  \caption{{\label{tab:4}}Redox solvation free energy $\Delta \mu_s$  and
    redox entropy $\Delta s_s$  in for the Red/Ox states of
    PC. Also listed are the reorganization energy and
    reorganization entropy, $s_{\lambda}=-\partial \lambda/ \partial T$. All energies
    are in eV and entropies are in meV/K, $T=300$ K. 
  }
\begin{ruledtabular}
\begin{tabular}{lllll}
Method & $\Delta \mu_s$ & $\Delta s_s$ & $\lambda_s$ & $s_{\lambda}$ \\
\hline
DelPhi with vdW cavity  &  $-9.92$ & $-1.25$ &  3.62 & 0.97 \\
DelPhi with solvent-accessible cavity  & $-7.12$ & $-0.45$ & 0.64 & 0.005\\
Non-local polarization response functions\footnotemark[1]  & $-6.01(-1.33)$ & $-10.5(4.8)$ & 1.30(0.55) &
$3.2(2.0)$\\
Molecular Dynamics\footnotemark[2] & $-2.81$  & $-7.40$ & 0.54 &
$10.7$ \\
Experiment & & $-0.4$\footnotemark[3]  &  &  \\
           & & $-1.4$\footnotemark[4]  &  &  \\
\end{tabular}
\end{ruledtabular}
\footnotetext[1]{The numbers in the parentheses indicate 
  contributions to the solvation free energy and entropy from density fluctuations.}  
\footnotetext[2]{MD results refer to linear fits through the
  simulation points.}
\footnotetext[3]{Ref.\ \onlinecite{Sola:99}.}
\footnotetext[4]{ Ref.\ \onlinecite{Sailasuta:79}.}
\end{table*}

Despite the right sign of the redox entropy, the magnitudes of both
the Gibbs solvation energy and the entropy are markedly different in
continuum and microscopic/simulation approaches: $\Delta\mu_s$ is higher in
the standard implementation of DelPhi (vdW cavity) than the NRFT value
by a factor of 1.5 while the redox entropy is lower by a factor of
ten. The use of the solvent-accessible cavity brings the value of
$\Delta\mu_s$ in a closer proximity to the NRFT, but the redox entropy is
lowered even more (Table \ref{tab:4}). The magnitude of $\Delta\mu_s$ from
NRFT is significantly higher than from MD even if the density
component is subtracted from the total response. This initially comes
a bit of surprise given a good agreement between the polarization-NRFT
and MD values of $\langle V_{0s}\rangle_{\text{Ox}}$ reported in Table \ref{tab:2}
and Fig.\ \ref{fig:7}. We do not currently have a good explanation of
this disagreement (see Discussion below).

\begin{figure}
  \centering
  \includegraphics*[width=6.5cm]{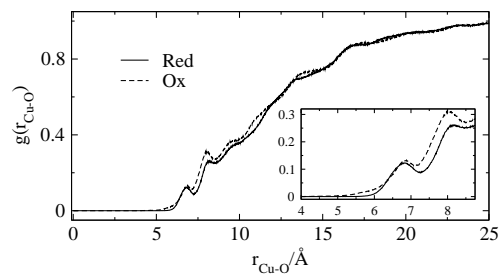}
  \caption{Pair distribution function between oxygen of
    water and Cu of PC in the reduced (Red) and oxidized (Ox) sates. }
  \label{fig:11}
\end{figure}

Concerning the reorganization energy calculations, the standard
DelPhi/vdW algorithm gives $\lambda_s$ three times larger than NRFT and
almost an order of magnitude larger than the MD simulations (Table
\ref{tab:4}).  The origin of large $\lambda_s$ in the standard (vdW)
implementation of the continuum model is in placing highly polar
dielectric into the small pocket near copper which water molecules do
not visit in MD simulations.  Figure \ref{fig:11} shows the pair
distribution function between the Cu ion and oxygen of water
testifying to the fact that water never comes to Cu closer than 6 \AA{}
and the maximum of the first solvation shell appears at 6.7 \AA.  The
addition of the solvent radius to the cavity corrects this error
reducing the reorganization entropy to the level of $0.64$ eV
consistent with the polarization component of the NRFT (0.75 eV, Table
\ref{tab:4}).

The combination of the absolute values of the reorganization energy
with the reorganization entropies clearly indicates that re-scaling of
the dielectric cavity, often employed in various continuum
formulations, does not solve the solvation problem. Entropies
calculated by including the effect of density fluctuations on the
solvent response are generally in better agreement with MD simulations
than the results obtained from polarization response only. All
solvation free energies obtained from such calculations, however,
significantly exceed the simulation results. The problem lies in
the assumed dense structure of the liquid around the solute, which is
obviously not realized for hydrated protein. Capturing density
fluctuations is essential for the correct calculations of the
entropies, but the effective density should be adjusted to that of the
structurally loose hydration layer at the protein surface.

\begin{figure}
  \centering
  \includegraphics*[width=6.5cm]{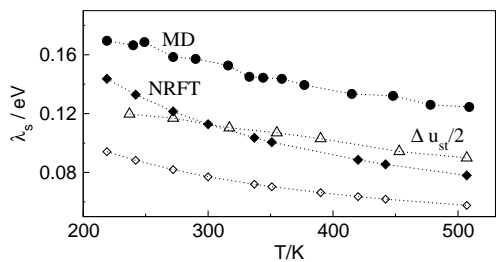}
  \caption{Reorganization energy of charge-transfer transition in
    $p$-nitroaniline dissolved in SPC/E water. The results are
    obtained by MD simulations\cite{DMjpca1:06} (closed circles) and
    NRFT calculations (diamonds). Closed and open diamonds refer to
    the full and polarization response, respectively. Triangles denote
    half of the Stokes shift from MD simulations; in the LRA, $\Delta
    u_{\text{st}}/2=\lambda_s$. The dielectric constants of SPC/E water at
    different temperatures, required for the NRFT input, were taken
    from MD simulations.\cite{DMjpca1:06}}
  \label{fig:12}
\end{figure}

One might argue that the failure to reproduce the results of the
simulations by the NRFT calculations can be traced back to the
deficiencies of the model and not to the specific structure created by
the protein in water. We believe that this objection cannot explain
the differences between our calculations and MD simulations. In order
to illustrate this point, we show in Fig.\ \ref{fig:12} the comparison
of the NRFT calculations with our recent MD results for solvation of a
small charge-transfer molecule $p$-nitroaniline in SPC/E
water.\cite{DMjpca1:06} The partial charges for this molecule, used in
the calculations were tabulated in Ref.\ \onlinecite{DMjpcb1:06}. The
comparison with the NRFT method is somewhat complicated in this case
by nonlinear solvation effects seen in the deviation of the half of
the Stokes shift from the reorganization energy. However, the NRFT
calculation with the density component included is in reasonable
agreement with the simulation data for both the reorganization energy
and entropy, while the polarization response alone clearly
underestimates the reorganization energy.

\section{Discussion}
\label{sec:6}
The formalism developed here is based on the LRA suggesting that
knowing the variance of electrostatic fluctuations around a neutral
repulsive solute is sufficient to calculate the solvation chemical
potential [Eq.\ (\ref{eq:1-2})].  One can arrive at Eq.\
\eqref{eq:1-2} from the following simple considerations. The chemical
potential of solvation can be written as the integral over the
magnitude of the solute-solvent electrostatic interaction $\epsilon=V_{0s}$
as follows
\begin{equation}
  \label{eq:6-1}
  e^{-\beta \mu_{0s}} = \int d\epsilon P(\epsilon,\beta) e^{-\beta \epsilon} .
\end{equation}
Here the probability density of reaching the value $\epsilon$ is obtained by
taking the statistical average over the reference Hamiltonian $H_0$
which includes all the interaction potentials in the system except the
solute-solvent electrostatic potential $V_{0s}$:
\begin{equation}
  \label{eq:6-2}
  P(\epsilon,\beta) = Q_0^{-1} \int \delta \left(\epsilon - V_{0s} \right) e^{-\beta H_0} d\Gamma ,
\end{equation}
where $Q_0=\int \exp[-\beta H_0] d\Gamma$ and $\Gamma$ denotes the phase space
volume.\cite{Landau5} Since the solute-solvent component of $H_0$
includes mostly isotropic short-range interactions, one can put $\langle
V_{0s}\rangle_0=0$. In addition, the Gaussian approximation can be applied to
$P(\epsilon,\beta)$
\begin{equation}
  \label{eq:6-3}
  P(\epsilon,\beta) \propto \exp\left[-\frac{\epsilon^2}{2 \sigma_{\epsilon}^2 }  \right] 
\end{equation}
with
\begin{equation}
  \label{eq:6-3-1}
  \sigma_{\epsilon}^2 = \langle (\delta V_{0s})^2\rangle_0 .
\end{equation}
Combining Eqs.\ \eqref{eq:6-1} and \eqref{eq:6-3}, one immediately
arrives at Eq.\ \eqref{eq:1-2}. 

One of the important lessons of Eqs.\ \eqref{eq:6-1} and
\eqref{eq:6-2} is that all the thermodynamic information required for
the solvation problem is contained in the distribution of
electrostatic potential fluctuations around a ``non-polar'' solute in
which all the electrostatic interactions with the solvent have been
switched off. For large solutes, the spectrum of electrostatic
fluctuations around the repulsive solute core will ultimately
determine the thermodynamics of solvation. It is important to stress
that Eqs.\ \eqref{eq:6-1} and \eqref{eq:6-2} are exact and this
statement is not limited to linear solvation only.

The reasoning outlined above works well for small and medium-size
solutes, as is exemplified in Fig.\ \ref{fig:12}. However, the
application of the same procedure to protein solvation studied in
this paper has encountered some difficulties suggesting that methods
developed over several decades and successfully applied to solvation
of small solutes in dense polar solvents are probably not directly
transferable without significant modifications to mesoscopic hydration
of proteins.  The application of our algorithm to the calculation of
the average solute-solvent interaction energy turned out to be quite
successful when the density profile of water around the protein is
approximated by a step function (Fig.\ \ref{fig:7}). The calculations
agree with MD within simulation uncertainties ($\leq 5$\%).  As mentioned
above, this comes as a result of averaging between tight and loose
water structures at the protein surface.  The application of the same
procedure to solvation of the difference charges of the active site
(reorganization energy, Fig.\ \ref{fig:10}) was less successful, but
the agreement is probably still acceptable, in particular at lowest
temperatures. 

Where the calculations and simulations come in significant
disagreement is for the free energy of solvation obtained in
simulations as the mean of two vertical transition energies [Eq.\
(\ref{eq:5-12})]. Some recent simulations of cavities in force-field
water models\cite{Ashbaugh:00,Rajamani:04} and of uncharged
protein\cite{Cerutti:07} have suggested a possible origin of this
effect.  It was found that water structured at the protein surface
creates a positive potential within an uncharged cavity/protein. In
terms of Eq.\ (\ref{eq:6-3}) this implies a constant shift of the
solute-solvent energy $\epsilon \to \epsilon-\epsilon_0$. Since we have assumed random
orientations of water around an uncharged solute, as is indeed the case
for small solutes,\cite{DMjpca1:06} our calculations do not include
the effect of a positive background potential and include only 
changes of the potential in response to protein's charges. In the case
of a constant background potential $\Phi$ Eq.\ (\ref{eq:5-12}) modifies
to
\begin{equation}
  \label{eq:6-3-2}
  \Delta \mu_s = \left(\langle\Delta V_{0s} \rangle_{\text{Ox}} + \langle\Delta V_{0s}
    \rangle_{\text{Red}}\right)/2 - \Delta q \Phi ,  
\end{equation}
where $\Delta q = q_{\text{Red}}-q_{\text{Ox}} = -1$.
Ashbaugh\cite{Ashbaugh:00} reported an average positive potential of
about $e \Phi \simeq 9$ kcal/mol for cavities in SPC water comparable in size
with PC. In terms of our calculations, it amounts a positive shift of
the simulation data by about 0.4 eV which will increase the current
gap of about 3.2 eV (300 K) between the MD and NRFT.

The origin in the difference in solvation free energies between
calculations and MD simulations might be related to the weakly
dewetted water density profile near the active site. This would imply
that some of the properties of water structure around large cavities
expelled by the protein from its volume are quite different from the
common experience gained with small solutes. This qualitative
difference between small-size and large-size solvation has recently
gained appreciation for hydrophobic solvation\cite{ChandlerNature:05}
as we discuss next.

The Gaussian model of hydrophobic solvation goes back to the
Pratt-Chandler theory of hydrophobicity\cite{PrattC:77} recently
extended by Pratt and co-workers.\cite{GardePRL:96,Ashbaugh:06} The
formulation of the theory of hydrophobic solvation follows a path
similar to the one outlined in Eqs.\ (\ref{eq:6-1})--(\ref{eq:6-3-1})
asking what is the free energy $\mu_{\Omega}$ needed to solvate a solute of
volume $\Omega_0$. It is given by the Gaussian
probability\cite{GardePRL:96,ChandlerNature:05}
\begin{equation}
  \label{eq:6-4}
  \beta \mu_{\Omega} \simeq \frac{\rho^2 \Omega_0^2}{2\chi_{\Omega}}
\end{equation}
with the fluctuation of the number of solvent particles in volume $\Omega_0$
\begin{equation}
  \label{eq:6-5}
 \chi_{\Omega} = \langle (\delta N)^2 \rangle_{\Omega} = \rho\Omega_0 + \rho^2 \int_{\Omega_0}
  d\mathbf{r} d\mathbf{r}' h_{ss}( | \mathbf{r} - \mathbf{r}' | ) .  
\end{equation}
In Eq.\ (\ref{eq:6-5}), $h_{ss}(r)$ is the pair correlation function
of the homogeneous solvent.

The Gaussian probability of electron transfer carries a close
similarity with Eq.\ (\ref{eq:6-4}) giving the activation free energy
as
\begin{equation}
  \label{eq:6-6}
  \beta \mu^{\text{act}} = \frac{X_0^2}{2 \sigma_{\epsilon}^2} ,
\end{equation}
where $X_0$ is the average vertical donor-acceptor energy gap 
and $\sigma_{\epsilon}^2$ is the variance of the solute-solvent
interaction potential when $V_{0s}=\Delta V_{0s}$ is used in Eqs.\
(\ref{eq:6-2}) and (\ref{eq:6-3}). Not surprisingly, the structure of 
the equation for $\sigma_{\epsilon}^2$ resembles Eq.\ (\ref{eq:6-5}):
\begin{equation}
  \label{eq:6-7}
  \begin{split}
  \sigma_{\epsilon}^2 &= \rho \int V_{0s}^2(1) d\Gamma_1 \\
                 & + \rho^2 \int
  V_{0s}(1)V_{0s}(2)h_{ss}(1,2) d\Gamma_1 d\Gamma_2 \\
                 & + \rho^2 \int V_{0s}^2(1) h_{ss}(1,2) \theta_0(2) d\Gamma_1  d\Gamma_2 .
 \end{split}    
\end{equation}
The last summand in this equation represents the density component of
the response since it transforms into the $\mathbf{k}$-space integral
with the density structure factor of the liquid [Eq.\
(\ref{eq:1-15-1})]. 

Equation (\ref{eq:6-7}) carries a close resemblance with the
Pratt-Chandler theory of hydrophobicity.\cite{PrattC:77} The sum of
the first and the third terms is the average of the squared
solute-solvent potential $V_{0s}^2(1)$ over the solute-solvent density
profile\cite{Lum:99}
\begin{equation}
  \label{eq:6-8}
  \rho(\mathbf{r}) = \rho (1+c_{0s}(\mathbf{r})) + 
          \rho^2 \int c_{0s}(\mathbf{r'}) h_{ss}( |\mathbf{r}-\mathbf{r}' | )d\mathbf{r}' 
\end{equation}
in which the solute-solvent direct correlation function
$c_{0s}(\mathbf{r})$ is replaced by its lowest density
expansion,\cite{Hansen:03} $c_{0s}(\mathbf{r})\simeq - \theta_0(\mathbf{r})$
($\theta_0(\mathbf{r})$ is one inside the solute and zero otherwise).  The
observation that the density profile around solutes of size $\geq 1$ nm
can be approximated by a step function (Fig.\ \ref{fig:11}, also see
Fig.\ 3 in Ref.\ \onlinecite{ChandlerNature:05}) amounts to neglecting
the second term in Eq.\ (\ref{eq:6-8}) and, correspondingly, the
density components $\bm{\chi}_d$ in the solvent response function [Eq.\
(\ref{eq:1-6-2})].

The fact that the spectrum of electrostatic potential fluctuations
around a non-polar solute gives complete information about polar
solvation brings this latter problem in close relation to the problem
of hydrophobic solvation. It was realized in recent years that
hydrophobic solvation of small and large solutes are qualitatively
different.\cite{Lum:99,ChandlerNature:05,Choudhury:07} Solvation
character changes at the critical size of $\simeq 1$ nm from
entropy-dominated solvation of small solutes (Gaussian
statistics\cite{GardePRL:96}) to enthalpy-dominated solvation of large
solutes driven by the creation of the solute-solvent interface
(non-Gaussian statistics\cite{Huang:00}).  The solvent interface
around large solutes involves partial dewetting, strongly sensitive to
the strength of solute-solvent attractions\cite{ZhouScience:04} and
the appearance of large-size interfacial density
fluctuations.\cite{ChandlerNature:05} In case of protein solvation,
surface water creates a nonzero potential around uncharged proteins
discussed above, while density fluctuations lead to complex
protein-solvent dynamics\cite{Tarek_PhysRevLett:02} and are probably
connected to ``slaving'' of the protein dynamics by the
solvent.\cite{Fenimore:04}

These new features observed for hydrophobic solvation at the
nano-scale will affect the thermodynamics of polar solvation.  It is
currently not clear how the cross-over from the Gaussian regime of
small solutes to the interface-dominated regime of large solutes will
translate to the problem of electrostatic solvation. The calculations
and simulations performed in this study gave some insights into the
kind of problems which the theory needs to address. It is clear that
the inhomogeneous nature of the protein surface requires algorithms
for the local density profile\cite{Weeks:02} or the solvent-accessible
surface\cite{Dzubiella:06} to be a part of a qualitative solvation
theory.  The current formulation can be improved by using a local
density approximation as, for instance, applied by Ramirez and
Borgis.\cite{RamirezJPCB:05} In this approach, the bulk dipolar
density $y$ [Eq.\ (\ref{eq:1-6})] is replaced by the local dipolar
density $y(\mathbf{r}) = (4\pi/9)\beta m'^2 \rho(\mathbf{r}) +
(4\pi/3)\alpha\rho(\mathbf{r})$ with the local density profile calculated from,
for instance, the Lum-Chandler-Weeks theory.\cite{Lum:99} Our
computational algorithm will then require the Fourier transform of the
field $\mathbf{E}_0(\mathbf{r})\rho(\mathbf{r})/ \rho$ instead of
$\mathbf{E}_0(\mathbf{r})(1-\theta_0(\mathbf{r}))$ in the current
implementation.  In addition, density fluctuations of the interfacial
region need to be addressed since the mean position of the interface
has no physical significance in the presence of large-amplitude
fluctuations.\cite{ChandlerNature:05} What is also clear is that the
density fluctuations of the interfacial region present a nuclear mode,
largely diminished for solvation of small solutes, which grows in
significance for solvation of biopolymers.  Future theoretical
development needs to address this physical reality.

Both NRFT and MD simulations predict a substantial temperature
dependence of the redox potential (ca.\ $\simeq 5-7$ mV/K). This magnitude
of the temperature variation is prohibitively high since a temperature
change of $\simeq 15$ K would shift the redox potential by about 100 mV
potentially terminating many enzymetic redox reactions.  The
theoretically predicted redox entropies refer to zero ionic strength
and the solvent contribution to the redox potential. The contribution
from the protein to the overall redox potential is available from our
simulations, but it depends weakly on temperature contributing only $\simeq
-0.9$ mV/K to the redox entropy. The combined solvent/protein entropy
is then significantly higher than redox entropies experimentally
observed in buffered solutions\cite{Sailasuta:79,Sola:99} (Table
\ref{tab:4}). This difference raises the question of the role of the
ionic atmosphere in stabilizing the redox potential of
mettalloproteins. The existing experimental evidence for small redox
molecules\cite{Hupp:93,Dalessandro:05} and simulations of
metalloproteins\cite{Swartz:96} all indicate a substantial effect of
the ionic atmosphere on the redox entropy, which might compensate for
the large entropy due to water and protein. Since simulations have
little to say about the ionic strength effects, laboratory
measurements of redox entropy at different buffer concentrations are
required to shed more light on this problem.

\begin{acknowledgments}
  This research was supported by the National Science Foundation
  (CHE-0616646). The code used for the NRFT calculations is available
  at \verb|http://theochemlab.asu.edu/codes.html|. We are
  grateful to Marco Sola for introducing us to the problem of protein
  redox entropy.
\end{acknowledgments}

\bibliography{/home/dmitry/p/bib/chem_abbr,/home/dmitry/p/bib/photosynthNew,/home/dmitry/p/bib/et,/home/dmitry/p/bib/dm,/home/dmitry/p/bib/protein,/home/dmitry/p/bib/solvation,/home/dmitry/p/bib/bioet,/home/dmitry/p/bib/etnonlin,/home/dmitry/p/bib/dynamics,/home/dmitry/p/bib/dielectric,/home/dmitry/p/bib/liquids,/home/dmitry/p/bib/simulations,/home/dmitry/p/bib/heteroet,/home/dmitry/p/bib/glass}

\begin{thebibliography}{111}
\expandafter\ifx\csname natexlab\endcsname\relax\def\natexlab#1{#1}\fi
\expandafter\ifx\csname bibnamefont\endcsname\relax
  \def\bibnamefont#1{#1}\fi
\expandafter\ifx\csname bibfnamefont\endcsname\relax
  \def\bibfnamefont#1{#1}\fi
\expandafter\ifx\csname citenamefont\endcsname\relax
  \def\citenamefont#1{#1}\fi
\expandafter\ifx\csname url\endcsname\relax
  \def\url#1{\texttt{#1}}\fi
\expandafter\ifx\csname urlprefix\endcsname\relax\def\urlprefix{URL }\fi
\providecommand{\bibinfo}[2]{#2}
\providecommand{\eprint}[2][]{\url{#2}}

\bibitem[{\citenamefont{Brooks et~al.}(1988)\citenamefont{Brooks, Karplus, and
  Pettitt}}]{Brooks:88}
\bibinfo{author}{\bibfnamefont{C.~L.} \bibnamefont{Brooks}},
  \bibinfo{author}{\bibfnamefont{M.}~\bibnamefont{Karplus}}, \bibnamefont{and}
  \bibinfo{author}{\bibfnamefont{B.~M.} \bibnamefont{Pettitt}},
  \bibinfo{journal}{Adv. Chem. Phys.} \textbf{\bibinfo{volume}{71}},
  \bibinfo{pages}{1} (\bibinfo{year}{1988}).

\bibitem[{\citenamefont{Eisenberg and McLachlan}(1986)}]{Eisenberg:86}
\bibinfo{author}{\bibfnamefont{D.}~\bibnamefont{Eisenberg}} \bibnamefont{and}
  \bibinfo{author}{\bibfnamefont{A.~D.} \bibnamefont{McLachlan}},
  \bibinfo{journal}{Nature} \textbf{\bibinfo{volume}{319}},
  \bibinfo{pages}{199} (\bibinfo{year}{1986}).

\bibitem[{\citenamefont{Zhou et~al.}(2004{\natexlab{a}})\citenamefont{Zhou,
  Krilov, and Berne}}]{Zhou:04}
\bibinfo{author}{\bibfnamefont{R.}~\bibnamefont{Zhou}},
  \bibinfo{author}{\bibfnamefont{G.}~\bibnamefont{Krilov}}, \bibnamefont{and}
  \bibinfo{author}{\bibfnamefont{B.~J.} \bibnamefont{Berne}},
  \bibinfo{journal}{J. Phys. Chem. B} \textbf{\bibinfo{volume}{108}},
  \bibinfo{pages}{7528} (\bibinfo{year}{2004}{\natexlab{a}}).

\bibitem[{\citenamefont{Mueller et~al.}(2006)\citenamefont{Mueller, Katsov, and
  Schick}}]{Mueller:06}
\bibinfo{author}{\bibfnamefont{M.}~\bibnamefont{Mueller}},
  \bibinfo{author}{\bibfnamefont{K.}~\bibnamefont{Katsov}}, \bibnamefont{and}
  \bibinfo{author}{\bibfnamefont{M.}~\bibnamefont{Schick}},
  \bibinfo{journal}{Phys. Rep.} \textbf{\bibinfo{volume}{434}},
  \bibinfo{pages}{113} (\bibinfo{year}{2006}).

\bibitem[{\citenamefont{Gohlke and Thorpe}(2006)}]{Gohlke:06}
\bibinfo{author}{\bibfnamefont{H.}~\bibnamefont{Gohlke}} \bibnamefont{and}
  \bibinfo{author}{\bibfnamefont{M.~F.} \bibnamefont{Thorpe}},
  \bibinfo{journal}{Biophys. J.} \textbf{\bibinfo{volume}{91}},
  \bibinfo{pages}{2115} (\bibinfo{year}{2006}).

\bibitem[{\citenamefont{Rocchia et~al.}(2002)\citenamefont{Rocchia, Sridharan,
  Nicholls, Alexov, Chiabrera, and Honig}}]{delphi02}
\bibinfo{author}{\bibfnamefont{W.}~\bibnamefont{Rocchia}},
  \bibinfo{author}{\bibfnamefont{S.}~\bibnamefont{Sridharan}},
  \bibinfo{author}{\bibfnamefont{A.}~\bibnamefont{Nicholls}},
  \bibinfo{author}{\bibfnamefont{E.}~\bibnamefont{Alexov}},
  \bibinfo{author}{\bibfnamefont{A.}~\bibnamefont{Chiabrera}},
  \bibnamefont{and} \bibinfo{author}{\bibfnamefont{B.}~\bibnamefont{Honig}},
  \bibinfo{journal}{J.\ Comp.\ Chem.} \textbf{\bibinfo{volume}{23}},
  \bibinfo{pages}{128} (\bibinfo{year}{2002}).

\bibitem[{\citenamefont{Schaefer and Karplus}(1996)}]{Schaefer:96}
\bibinfo{author}{\bibfnamefont{M.}~\bibnamefont{Schaefer}} \bibnamefont{and}
  \bibinfo{author}{\bibfnamefont{M.}~\bibnamefont{Karplus}},
  \bibinfo{journal}{J. Phys. Chem.} \textbf{\bibinfo{volume}{100}},
  \bibinfo{pages}{1578} (\bibinfo{year}{1996}).

\bibitem[{\citenamefont{Gunner and Honig}(1991)}]{Gunner:91}
\bibinfo{author}{\bibfnamefont{M.~R.} \bibnamefont{Gunner}} \bibnamefont{and}
  \bibinfo{author}{\bibfnamefont{B.}~\bibnamefont{Honig}},
  \bibinfo{journal}{Proc. Natl. Acad. Sci.} \textbf{\bibinfo{volume}{88}},
  \bibinfo{pages}{9151} (\bibinfo{year}{1991}).

\bibitem[{\citenamefont{Sharp}(1998)}]{Sharp:98}
\bibinfo{author}{\bibfnamefont{K.~A.} \bibnamefont{Sharp}},
  \bibinfo{journal}{Biophys. J.} \textbf{\bibinfo{volume}{73}},
  \bibinfo{pages}{1241} (\bibinfo{year}{1998}).

\bibitem[{\citenamefont{Siriwong et~al.}(2003)\citenamefont{Siriwong, Voityuk,
  Newton, and R{\"o}sch}}]{Siriwong:03}
\bibinfo{author}{\bibfnamefont{K.}~\bibnamefont{Siriwong}},
  \bibinfo{author}{\bibfnamefont{A.~A.} \bibnamefont{Voityuk}},
  \bibinfo{author}{\bibfnamefont{M.~D.} \bibnamefont{Newton}},
  \bibnamefont{and}
  \bibinfo{author}{\bibfnamefont{N.}~\bibnamefont{R{\"o}sch}},
  \bibinfo{journal}{J. Phys. Chem. B} \textbf{\bibinfo{volume}{107}},
  \bibinfo{pages}{2595} (\bibinfo{year}{2003}).

\bibitem[{\citenamefont{Qiu et~al.}(1997)\citenamefont{Qiu, Shenkin, Hollinger,
  and Still}}]{Qiu:97}
\bibinfo{author}{\bibfnamefont{D.}~\bibnamefont{Qiu}},
  \bibinfo{author}{\bibfnamefont{P.}~\bibnamefont{Shenkin}},
  \bibinfo{author}{\bibfnamefont{F.}~\bibnamefont{Hollinger}},
  \bibnamefont{and} \bibinfo{author}{\bibfnamefont{W.}~\bibnamefont{Still}},
  \bibinfo{journal}{J.\ Phys.\ Chem.\ A} \textbf{\bibinfo{volume}{101}},
  \bibinfo{pages}{3005} (\bibinfo{year}{1997}).

\bibitem[{\citenamefont{Vath et~al.}(1999)\citenamefont{Vath, Zimmt, Matyushov,
  and Voth}}]{DMjpcb:99}
\bibinfo{author}{\bibfnamefont{P.}~\bibnamefont{Vath}},
  \bibinfo{author}{\bibfnamefont{M.~B.} \bibnamefont{Zimmt}},
  \bibinfo{author}{\bibfnamefont{D.~V.} \bibnamefont{Matyushov}},
  \bibnamefont{and} \bibinfo{author}{\bibfnamefont{G.~A.} \bibnamefont{Voth}},
  \bibinfo{journal}{J. Phys. Chem. B} \textbf{\bibinfo{volume}{103}},
  \bibinfo{pages}{9130} (\bibinfo{year}{1999}).

\bibitem[{\citenamefont{Matyushov}(2004{\natexlab{a}})}]{DMjcp2:04}
\bibinfo{author}{\bibfnamefont{D.~V.} \bibnamefont{Matyushov}},
  \bibinfo{journal}{J. Chem. Phys.} \textbf{\bibinfo{volume}{120}},
  \bibinfo{pages}{7532} (\bibinfo{year}{2004}{\natexlab{a}}).

\bibitem[{\citenamefont{Milischuk et~al.}(2006)\citenamefont{Milischuk,
  Matyushov, and Newton}}]{DMcp:06}
\bibinfo{author}{\bibfnamefont{A.~A.} \bibnamefont{Milischuk}},
  \bibinfo{author}{\bibfnamefont{D.~V.} \bibnamefont{Matyushov}},
  \bibnamefont{and} \bibinfo{author}{\bibfnamefont{M.~D.}
  \bibnamefont{Newton}}, \bibinfo{journal}{Chem. Phys.}
  \textbf{\bibinfo{volume}{324}}, \bibinfo{pages}{172} (\bibinfo{year}{2006}).

\bibitem[{\citenamefont{Ghorai and Matyushov}(2006{\natexlab{a}})}]{DMjpca1:06}
\bibinfo{author}{\bibfnamefont{P.~K.} \bibnamefont{Ghorai}} \bibnamefont{and}
  \bibinfo{author}{\bibfnamefont{D.~V.} \bibnamefont{Matyushov}},
  \bibinfo{journal}{J. Phys. Chem. A} \textbf{\bibinfo{volume}{110}},
  \bibinfo{pages}{8857} (\bibinfo{year}{2006}{\natexlab{a}}).

\bibitem[{\citenamefont{Zhou et~al.}(2004{\natexlab{b}})\citenamefont{Zhou,
  Huang, Margulis, and Berne}}]{ZhouScience:04}
\bibinfo{author}{\bibfnamefont{R.}~\bibnamefont{Zhou}},
  \bibinfo{author}{\bibfnamefont{X.}~\bibnamefont{Huang}},
  \bibinfo{author}{\bibfnamefont{C.~J.} \bibnamefont{Margulis}},
  \bibnamefont{and} \bibinfo{author}{\bibfnamefont{B.~J.} \bibnamefont{Berne}},
  \bibinfo{journal}{Science} \textbf{\bibinfo{volume}{305}},
  \bibinfo{pages}{1605} (\bibinfo{year}{2004}{\natexlab{b}}).

\bibitem[{\citenamefont{Tarek and Tobias}(2002)}]{Tarek_PhysRevLett:02}
\bibinfo{author}{\bibfnamefont{M.}~\bibnamefont{Tarek}} \bibnamefont{and}
  \bibinfo{author}{\bibfnamefont{D.~J.} \bibnamefont{Tobias}},
  \bibinfo{journal}{Phys. Rev. Lett.} \textbf{\bibinfo{volume}{88}},
  \bibinfo{pages}{138101} (\bibinfo{year}{2002}).

\bibitem[{\citenamefont{Choudhury and Pettitt}(2007)}]{Choudhury:07}
\bibinfo{author}{\bibfnamefont{N.}~\bibnamefont{Choudhury}} \bibnamefont{and}
  \bibinfo{author}{\bibfnamefont{B.}~\bibnamefont{Pettitt}},
  \bibinfo{journal}{J.\ Am.\ Chem.\ Soc.} \textbf{\bibinfo{volume}{129}},
  \bibinfo{pages}{4847} (\bibinfo{year}{2007}).

\bibitem[{\citenamefont{Matyushov}(1993)}]{DMcp:93}
\bibinfo{author}{\bibfnamefont{D.~V.} \bibnamefont{Matyushov}},
  \bibinfo{journal}{Chem. Phys.} \textbf{\bibinfo{volume}{174}},
  \bibinfo{pages}{199} (\bibinfo{year}{1993}).

\bibitem[{\citenamefont{Ubbink et~al.}(1998)\citenamefont{Ubbink, Ejdeb{\"a}ck,
  Karlsson, and Bendall}}]{Ubbink:98}
\bibinfo{author}{\bibfnamefont{M.}~\bibnamefont{Ubbink}},
  \bibinfo{author}{\bibfnamefont{M.}~\bibnamefont{Ejdeb{\"a}ck}},
  \bibinfo{author}{\bibfnamefont{B.~G.} \bibnamefont{Karlsson}},
  \bibnamefont{and} \bibinfo{author}{\bibfnamefont{D.~S.}
  \bibnamefont{Bendall}}, \bibinfo{journal}{Structure}
  \textbf{\bibinfo{volume}{6}}, \bibinfo{pages}{323} (\bibinfo{year}{1998}).

\bibitem[{\citenamefont{Sailasuta et~al.}(1979)\citenamefont{Sailasuta, Anson,
  and Gray}}]{Sailasuta:79}
\bibinfo{author}{\bibfnamefont{N.}~\bibnamefont{Sailasuta}},
  \bibinfo{author}{\bibfnamefont{F.~C.} \bibnamefont{Anson}}, \bibnamefont{and}
  \bibinfo{author}{\bibfnamefont{H.~B.} \bibnamefont{Gray}},
  \bibinfo{journal}{J. Am. Chem. Soc.} \textbf{\bibinfo{volume}{101}},
  \bibinfo{pages}{455} (\bibinfo{year}{1979}).

\bibitem[{\citenamefont{Battistuzzi et~al.}(1999)\citenamefont{Battistuzzi,
  Borsari, Loschi, Righi, and Sola}}]{Sola:99}
\bibinfo{author}{\bibfnamefont{G.}~\bibnamefont{Battistuzzi}},
  \bibinfo{author}{\bibfnamefont{M.}~\bibnamefont{Borsari}},
  \bibinfo{author}{\bibfnamefont{L.}~\bibnamefont{Loschi}},
  \bibinfo{author}{\bibfnamefont{F.}~\bibnamefont{Righi}}, \bibnamefont{and}
  \bibinfo{author}{\bibfnamefont{M.}~\bibnamefont{Sola}}, \bibinfo{journal}{J.
  Am. Chem. Soc.} \textbf{\bibinfo{volume}{121}}, \bibinfo{pages}{501}
  (\bibinfo{year}{1999}).

\bibitem[{\citenamefont{Battistuzzi et~al.}(2003)\citenamefont{Battistuzzi,
  Bellei, Borsari, Canters, de~Waal, Jeuken, Ranieri, and
  Sola}}]{Battistuzzi:03}
\bibinfo{author}{\bibfnamefont{G.}~\bibnamefont{Battistuzzi}},
  \bibinfo{author}{\bibfnamefont{M.}~\bibnamefont{Bellei}},
  \bibinfo{author}{\bibfnamefont{M.}~\bibnamefont{Borsari}},
  \bibinfo{author}{\bibfnamefont{G.~W.} \bibnamefont{Canters}},
  \bibinfo{author}{\bibfnamefont{E.}~\bibnamefont{de~Waal}},
  \bibinfo{author}{\bibfnamefont{L.~J.~C.} \bibnamefont{Jeuken}},
  \bibinfo{author}{\bibfnamefont{A.}~\bibnamefont{Ranieri}}, \bibnamefont{and}
  \bibinfo{author}{\bibfnamefont{M.}~\bibnamefont{Sola}},
  \bibinfo{journal}{Biochemistry} \textbf{\bibinfo{volume}{42}},
  \bibinfo{pages}{9214} (\bibinfo{year}{2003}).

\bibitem[{\citenamefont{Lockwood et~al.}(2001)\citenamefont{Lockwood, Cheng,
  and Rossky}}]{Lockwood:01}
\bibinfo{author}{\bibfnamefont{D.~M.} \bibnamefont{Lockwood}},
  \bibinfo{author}{\bibfnamefont{Y.-K.} \bibnamefont{Cheng}}, \bibnamefont{and}
  \bibinfo{author}{\bibfnamefont{P.~J.} \bibnamefont{Rossky}},
  \bibinfo{journal}{Chem. Phys. Lett.} \textbf{\bibinfo{volume}{345}},
  \bibinfo{pages}{159} (\bibinfo{year}{2001}).

\bibitem[{\citenamefont{Datta et~al.}(2004)\citenamefont{Datta, Sudhamsu, and
  Pandey}}]{Datta:04}
\bibinfo{author}{\bibfnamefont{S.~N.} \bibnamefont{Datta}},
  \bibinfo{author}{\bibfnamefont{J.}~\bibnamefont{Sudhamsu}}, \bibnamefont{and}
  \bibinfo{author}{\bibfnamefont{A.}~\bibnamefont{Pandey}},
  \bibinfo{journal}{J. Phys. Chem. B} \textbf{\bibinfo{volume}{108}},
  \bibinfo{pages}{8007} (\bibinfo{year}{2004}).

\bibitem[{\citenamefont{Hansen and Led}(2004)}]{Hansen:04}
\bibinfo{author}{\bibfnamefont{D.~F.} \bibnamefont{Hansen}} \bibnamefont{and}
  \bibinfo{author}{\bibfnamefont{J.~J.} \bibnamefont{Led}},
  \bibinfo{journal}{J. Am. Chem. Soc.} \textbf{\bibinfo{volume}{126}},
  \bibinfo{pages}{1247} (\bibinfo{year}{2004}).

\bibitem[{\citenamefont{Solomon et~al.}(2004)\citenamefont{Solomon, Szilagyi,
  DeBeerGeorge, and Basumallick}}]{Solomon:04}
\bibinfo{author}{\bibfnamefont{E.}~\bibnamefont{Solomon}},
  \bibinfo{author}{\bibfnamefont{R.}~\bibnamefont{Szilagyi}},
  \bibinfo{author}{\bibfnamefont{S.}~\bibnamefont{DeBeerGeorge}},
  \bibnamefont{and}
  \bibinfo{author}{\bibfnamefont{L.}~\bibnamefont{Basumallick}},
  \bibinfo{journal}{Chem. Rev.} \textbf{\bibinfo{volume}{104}},
  \bibinfo{pages}{419} (\bibinfo{year}{2004}).

\bibitem[{\citenamefont{Solomon}(2006)}]{Solomon:06}
\bibinfo{author}{\bibfnamefont{E.~I.} \bibnamefont{Solomon}},
  \bibinfo{journal}{Inorg. Chem.} \textbf{\bibinfo{volume}{45}},
  \bibinfo{pages}{8012} (\bibinfo{year}{2006}).

\bibitem[{\citenamefont{Guss et~al.}(1986)\citenamefont{Guss, Harrowell,
  Murata, Norris, and Freeman}}]{Guss:86}
\bibinfo{author}{\bibfnamefont{J.~M.} \bibnamefont{Guss}},
  \bibinfo{author}{\bibfnamefont{P.~R.} \bibnamefont{Harrowell}},
  \bibinfo{author}{\bibfnamefont{M.}~\bibnamefont{Murata}},
  \bibinfo{author}{\bibfnamefont{V.~A.} \bibnamefont{Norris}},
  \bibnamefont{and} \bibinfo{author}{\bibfnamefont{H.~C.}
  \bibnamefont{Freeman}}, \bibinfo{journal}{J. Mol. Biol.}
  \textbf{\bibinfo{volume}{192}}, \bibinfo{pages}{361} (\bibinfo{year}{1986}).

\bibitem[{\citenamefont{Stephens et~al.}(1996)\citenamefont{Stephens, Jollie,
  and Warshel}}]{Stephens:96}
\bibinfo{author}{\bibfnamefont{P.~J.} \bibnamefont{Stephens}},
  \bibinfo{author}{\bibfnamefont{D.~R.} \bibnamefont{Jollie}},
  \bibnamefont{and} \bibinfo{author}{\bibfnamefont{A.}~\bibnamefont{Warshel}},
  \bibinfo{journal}{Chem. Rev.} \textbf{\bibinfo{volume}{96}},
  \bibinfo{pages}{2491} (\bibinfo{year}{1996}).

\bibitem[{\citenamefont{Olsson et~al.}(2003)\citenamefont{Olsson, Hong, and
  Warshel}}]{Olsson:03}
\bibinfo{author}{\bibfnamefont{M.}~\bibnamefont{Olsson}},
  \bibinfo{author}{\bibfnamefont{G.}~\bibnamefont{Hong}}, \bibnamefont{and}
  \bibinfo{author}{\bibfnamefont{A.}~\bibnamefont{Warshel}},
  \bibinfo{journal}{J.\ Am.\ Chem.\ Soc.} \textbf{\bibinfo{volume}{125}},
  \bibinfo{pages}{5025} (\bibinfo{year}{2003}).

\bibitem[{\citenamefont{Yee et~al.}(1979)\citenamefont{Yee, Cave, Guyer, Tyma,
  and Weaver}}]{Yee:79}
\bibinfo{author}{\bibfnamefont{E.~L.} \bibnamefont{Yee}},
  \bibinfo{author}{\bibfnamefont{R.~J.} \bibnamefont{Cave}},
  \bibinfo{author}{\bibfnamefont{K.~L.} \bibnamefont{Guyer}},
  \bibinfo{author}{\bibfnamefont{P.~D.} \bibnamefont{Tyma}}, \bibnamefont{and}
  \bibinfo{author}{\bibfnamefont{M.~J.} \bibnamefont{Weaver}},
  \bibinfo{journal}{J. Am. Chem. Soc.} \textbf{\bibinfo{volume}{101}},
  \bibinfo{pages}{1131} (\bibinfo{year}{1979}).

\bibitem[{\citenamefont{Born}(1920)}]{Born:20}
\bibinfo{author}{\bibfnamefont{M.}~\bibnamefont{Born}}, \bibinfo{journal}{Z.
  Phys.} \textbf{\bibinfo{volume}{1}}, \bibinfo{pages}{45}
  (\bibinfo{year}{1920}).

\bibitem[{\citenamefont{Onsager}(1936)}]{Onsager:36}
\bibinfo{author}{\bibfnamefont{L.}~\bibnamefont{Onsager}}, \bibinfo{journal}{J.
  Am. Chem. Soc.} \textbf{\bibinfo{volume}{58}}, \bibinfo{pages}{1486}
  (\bibinfo{year}{1936}).

\bibitem[{\citenamefont{Kirkwood}(1934)}]{Kirkwood:34}
\bibinfo{author}{\bibfnamefont{J.~G.} \bibnamefont{Kirkwood}},
  \bibinfo{journal}{J. Chem. Phys.} \textbf{\bibinfo{volume}{2}},
  \bibinfo{pages}{351} (\bibinfo{year}{1934}).

\bibitem[{\citenamefont{Raineri and Friedman}(1999)}]{Raineri:99}
\bibinfo{author}{\bibfnamefont{F.~O.} \bibnamefont{Raineri}} \bibnamefont{and}
  \bibinfo{author}{\bibfnamefont{H.~L.} \bibnamefont{Friedman}},
  \bibinfo{journal}{Adv. Chem. Phys.} \textbf{\bibinfo{volume}{107}},
  \bibinfo{pages}{81} (\bibinfo{year}{1999}).

\bibitem[{\citenamefont{Yoshimori}(2004)}]{Yoshimori:04}
\bibinfo{author}{\bibfnamefont{A.}~\bibnamefont{Yoshimori}},
  \bibinfo{journal}{J. Theor. Comp. Chem.} \textbf{\bibinfo{volume}{3}},
  \bibinfo{pages}{117} (\bibinfo{year}{2004}).

\bibitem[{\citenamefont{Burghardta and Bagchi}(2006)}]{BagchiCP:06}
\bibinfo{author}{\bibfnamefont{I.}~\bibnamefont{Burghardta}} \bibnamefont{and}
  \bibinfo{author}{\bibfnamefont{B.}~\bibnamefont{Bagchi}},
  \bibinfo{journal}{Chem. Phys.} \textbf{\bibinfo{volume}{329}},
  \bibinfo{pages}{343} (\bibinfo{year}{2006}).

\bibitem[{\citenamefont{Biben et~al.}(1998)\citenamefont{Biben, Hansen, and
  Rosenfeld}}]{Biben:98}
\bibinfo{author}{\bibfnamefont{T.}~\bibnamefont{Biben}},
  \bibinfo{author}{\bibfnamefont{J.~P.} \bibnamefont{Hansen}},
  \bibnamefont{and}
  \bibinfo{author}{\bibfnamefont{Y.}~\bibnamefont{Rosenfeld}},
  \bibinfo{journal}{Phys. Rev. E} \textbf{\bibinfo{volume}{57}},
  \bibinfo{pages}{R3727} (\bibinfo{year}{1998}).

\bibitem[{\citenamefont{Ramirez et~al.}(2002)\citenamefont{Ramirez, Gebauer,
  Mareschal, and Borgis}}]{Ramirez:02}
\bibinfo{author}{\bibfnamefont{R.}~\bibnamefont{Ramirez}},
  \bibinfo{author}{\bibfnamefont{R.}~\bibnamefont{Gebauer}},
  \bibinfo{author}{\bibfnamefont{M.}~\bibnamefont{Mareschal}},
  \bibnamefont{and} \bibinfo{author}{\bibfnamefont{D.}~\bibnamefont{Borgis}},
  \bibinfo{journal}{Phys. Rev. E} \textbf{\bibinfo{volume}{66}},
  \bibinfo{pages}{031206} (\bibinfo{year}{2002}).

\bibitem[{\citenamefont{Ramirez and Borgis}(2005)}]{RamirezJPCB:05}
\bibinfo{author}{\bibfnamefont{R.}~\bibnamefont{Ramirez}} \bibnamefont{and}
  \bibinfo{author}{\bibfnamefont{D.}~\bibnamefont{Borgis}},
  \bibinfo{journal}{J.\ Phys.\ Chem.\ B} \textbf{\bibinfo{volume}{109}},
  \bibinfo{pages}{6754} (\bibinfo{year}{2005}).

\bibitem[{\citenamefont{Milischuk and Matyushov}(2006)}]{DMjcp3:06}
\bibinfo{author}{\bibfnamefont{A.~A.} \bibnamefont{Milischuk}}
  \bibnamefont{and} \bibinfo{author}{\bibfnamefont{D.~V.}
  \bibnamefont{Matyushov}}, \bibinfo{journal}{J. Chem. Phys.}
  \textbf{\bibinfo{volume}{124}}, \bibinfo{pages}{204502}
  (\bibinfo{year}{2006}).

\bibitem[{\citenamefont{LeBard et~al.}(2003)\citenamefont{LeBard, Lilichenko,
  Matyushov, Berlin, and Ratner}}]{DMjpcb1:03}
\bibinfo{author}{\bibfnamefont{D.~N.} \bibnamefont{LeBard}},
  \bibinfo{author}{\bibfnamefont{M.}~\bibnamefont{Lilichenko}},
  \bibinfo{author}{\bibfnamefont{D.~V.} \bibnamefont{Matyushov}},
  \bibinfo{author}{\bibfnamefont{Y.~A.} \bibnamefont{Berlin}},
  \bibnamefont{and} \bibinfo{author}{\bibfnamefont{M.~A.}
  \bibnamefont{Ratner}}, \bibinfo{journal}{J. Phys. Chem. B}
  \textbf{\bibinfo{volume}{107}}, \bibinfo{pages}{14509}
  (\bibinfo{year}{2003}).

\bibitem[{\citenamefont{Milischuk and Matyushov}(2005)}]{DMjcp4:05}
\bibinfo{author}{\bibfnamefont{A.~A.} \bibnamefont{Milischuk}}
  \bibnamefont{and} \bibinfo{author}{\bibfnamefont{D.~V.}
  \bibnamefont{Matyushov}}, \bibinfo{journal}{J. Chem. Phys.}
  \textbf{\bibinfo{volume}{123}}, \bibinfo{pages}{044501}
  (\bibinfo{year}{2005}).

\bibitem[{\citenamefont{Matyushov and Newton}(2001)}]{DMjpca:01}
\bibinfo{author}{\bibfnamefont{D.~V.} \bibnamefont{Matyushov}}
  \bibnamefont{and} \bibinfo{author}{\bibfnamefont{M.~D.}
  \bibnamefont{Newton}}, \bibinfo{journal}{J. Phys. Chem. A}
  \textbf{\bibinfo{volume}{105}}, \bibinfo{pages}{8516} (\bibinfo{year}{2001}).

\bibitem[{\citenamefont{Stell et~al.}(1981)\citenamefont{Stell, Patey, and
  H{{\o}}ye}}]{SPH:81}
\bibinfo{author}{\bibfnamefont{G.}~\bibnamefont{Stell}},
  \bibinfo{author}{\bibfnamefont{G.~N.} \bibnamefont{Patey}}, \bibnamefont{and}
  \bibinfo{author}{\bibfnamefont{J.~S.} \bibnamefont{H{{\o}}ye}},
  \bibinfo{journal}{Adv. Chem. Phys.} \textbf{\bibinfo{volume}{18}},
  \bibinfo{pages}{183} (\bibinfo{year}{1981}).

\bibitem[{\citenamefont{Ferenczy and Reynolds}(2001)}]{CReynolds:01}
\bibinfo{author}{\bibfnamefont{G.~G.} \bibnamefont{Ferenczy}} \bibnamefont{and}
  \bibinfo{author}{\bibfnamefont{C.~A.} \bibnamefont{Reynolds}},
  \bibinfo{journal}{J. Phys. Chem. A} \textbf{\bibinfo{volume}{105}},
  \bibinfo{pages}{11470} (\bibinfo{year}{2001}).

\bibitem[{\citenamefont{Matyushov and Ladanyi}(1999)}]{DMjcp1:99}
\bibinfo{author}{\bibfnamefont{D.~V.} \bibnamefont{Matyushov}}
  \bibnamefont{and} \bibinfo{author}{\bibfnamefont{B.~M.}
  \bibnamefont{Ladanyi}}, \bibinfo{journal}{J. Chem. Phys.}
  \textbf{\bibinfo{volume}{110}}, \bibinfo{pages}{994} (\bibinfo{year}{1999}).

\bibitem[{\citenamefont{Carter and Hynes}(1991)}]{Carter:91}
\bibinfo{author}{\bibfnamefont{E.~A.} \bibnamefont{Carter}} \bibnamefont{and}
  \bibinfo{author}{\bibfnamefont{J.~T.} \bibnamefont{Hynes}},
  \bibinfo{journal}{J. Chem. Phys.} \textbf{\bibinfo{volume}{94}},
  \bibinfo{pages}{5961} (\bibinfo{year}{1991}).

\bibitem[{\citenamefont{Andersen et~al.}(1976)\citenamefont{Andersen, Chandler,
  and Weeks}}]{WCA}
\bibinfo{author}{\bibfnamefont{H.~C.} \bibnamefont{Andersen}},
  \bibinfo{author}{\bibfnamefont{D.}~\bibnamefont{Chandler}}, \bibnamefont{and}
  \bibinfo{author}{\bibfnamefont{J.~D.} \bibnamefont{Weeks}},
  \bibinfo{journal}{Adv. Chem. Phys.} \textbf{\bibinfo{volume}{34}},
  \bibinfo{pages}{105} (\bibinfo{year}{1976}).

\bibitem[{\citenamefont{Hwang and Warshel}(1987)}]{Hwang:87}
\bibinfo{author}{\bibfnamefont{J.-K.} \bibnamefont{Hwang}} \bibnamefont{and}
  \bibinfo{author}{\bibfnamefont{A.}~\bibnamefont{Warshel}},
  \bibinfo{journal}{J. Am. Chem. Soc.} \textbf{\bibinfo{volume}{109}},
  \bibinfo{pages}{715} (\bibinfo{year}{1987}).

\bibitem[{\citenamefont{Kuharski et~al.}(1988)\citenamefont{Kuharski, Bader,
  Chandler, Sprik, Klein, and Impey}}]{Kuharski:88}
\bibinfo{author}{\bibfnamefont{R.~A.} \bibnamefont{Kuharski}},
  \bibinfo{author}{\bibfnamefont{J.~S.} \bibnamefont{Bader}},
  \bibinfo{author}{\bibfnamefont{D.}~\bibnamefont{Chandler}},
  \bibinfo{author}{\bibfnamefont{M.}~\bibnamefont{Sprik}},
  \bibinfo{author}{\bibfnamefont{M.~L.} \bibnamefont{Klein}}, \bibnamefont{and}
  \bibinfo{author}{\bibfnamefont{R.~W.} \bibnamefont{Impey}},
  \bibinfo{journal}{J. Chem. Phys.} \textbf{\bibinfo{volume}{89}},
  \bibinfo{pages}{3248} (\bibinfo{year}{1988}).

\bibitem[{\citenamefont{Blumberger and Sprik}(2005)}]{Blumberger:05}
\bibinfo{author}{\bibfnamefont{J.}~\bibnamefont{Blumberger}} \bibnamefont{and}
  \bibinfo{author}{\bibfnamefont{M.}~\bibnamefont{Sprik}}, \bibinfo{journal}{J.
  Phys. Chem. B} \textbf{\bibinfo{volume}{109}}, \bibinfo{pages}{6793}
  (\bibinfo{year}{2005}).

\bibitem[{\citenamefont{Fonseca and Ladanyi}(1994)}]{Fonseca:94}
\bibinfo{author}{\bibfnamefont{T.}~\bibnamefont{Fonseca}} \bibnamefont{and}
  \bibinfo{author}{\bibfnamefont{B.~M.} \bibnamefont{Ladanyi}},
  \bibinfo{journal}{J. Mol. Liq.} \textbf{\bibinfo{volume}{60}},
  \bibinfo{pages}{1} (\bibinfo{year}{1994}).

\bibitem[{\citenamefont{Lee and Hynes}(1988)}]{Lee:88}
\bibinfo{author}{\bibfnamefont{S.}~\bibnamefont{Lee}} \bibnamefont{and}
  \bibinfo{author}{\bibfnamefont{J.~T.} \bibnamefont{Hynes}},
  \bibinfo{journal}{J. Chem. Phys.} \textbf{\bibinfo{volume}{88}},
  \bibinfo{pages}{6853} (\bibinfo{year}{1988}).

\bibitem[{\citenamefont{Chandler}(1993)}]{Chandler:93}
\bibinfo{author}{\bibfnamefont{D.}~\bibnamefont{Chandler}},
  \bibinfo{journal}{Phys. Rev. E} \textbf{\bibinfo{volume}{48}},
  \bibinfo{pages}{2898} (\bibinfo{year}{1993}).

\bibitem[{\citenamefont{Hansen and McDonald}(2003)}]{Hansen:03}
\bibinfo{author}{\bibfnamefont{J.~P.} \bibnamefont{Hansen}} \bibnamefont{and}
  \bibinfo{author}{\bibfnamefont{I.~R.} \bibnamefont{McDonald}},
  \emph{\bibinfo{title}{Theory of Simple Liquids}}
  (\bibinfo{publisher}{Academic Press}, \bibinfo{address}{Amsterdam},
  \bibinfo{year}{2003}).

\bibitem[{\citenamefont{Weeks}(2002)}]{Weeks:02}
\bibinfo{author}{\bibfnamefont{J.~D.} \bibnamefont{Weeks}},
  \bibinfo{journal}{Annu. Rev. Phys. Chem.} \textbf{\bibinfo{volume}{53}},
  \bibinfo{pages}{533} (\bibinfo{year}{2002}).

\bibitem[{\citenamefont{Fried and Mukamel}(1990)}]{Fried:90}
\bibinfo{author}{\bibfnamefont{L.~E.} \bibnamefont{Fried}} \bibnamefont{and}
  \bibinfo{author}{\bibfnamefont{S.}~\bibnamefont{Mukamel}},
  \bibinfo{journal}{J. Chem. Phys.} \textbf{\bibinfo{volume}{93}},
  \bibinfo{pages}{932} (\bibinfo{year}{1990}).

\bibitem[{\citenamefont{Bagchi and Chandra}(1991)}]{Bagchi:91}
\bibinfo{author}{\bibfnamefont{B.}~\bibnamefont{Bagchi}} \bibnamefont{and}
  \bibinfo{author}{\bibfnamefont{A.}~\bibnamefont{Chandra}},
  \bibinfo{journal}{Adv. Chem. Phys.} \textbf{\bibinfo{volume}{80}},
  \bibinfo{pages}{1} (\bibinfo{year}{1991}).

\bibitem[{\citenamefont{Madden and Kivelson}(1984)}]{Madden:84}
\bibinfo{author}{\bibfnamefont{P.}~\bibnamefont{Madden}} \bibnamefont{and}
  \bibinfo{author}{\bibfnamefont{D.}~\bibnamefont{Kivelson}},
  \bibinfo{journal}{Adv. Chem. Phys.} \textbf{\bibinfo{volume}{56}},
  \bibinfo{pages}{467} (\bibinfo{year}{1984}).

\bibitem[{\citenamefont{B{{\"o}}ttcher}(1973)}]{Boettcher:73}
\bibinfo{author}{\bibfnamefont{C.~J.~F.} \bibnamefont{B{{\"o}}ttcher}},
  \emph{\bibinfo{title}{Theory of Electric Polarization}},
  vol.~\bibinfo{volume}{1} (\bibinfo{publisher}{Elsevier},
  \bibinfo{address}{Amsterdam}, \bibinfo{year}{1973}).

\bibitem[{\citenamefont{Kharkats et~al.}(1976)\citenamefont{Kharkats,
  Kornyshev, and Vorotyntsev}}]{KKV:76}
\bibinfo{author}{\bibfnamefont{Y.~I.} \bibnamefont{Kharkats}},
  \bibinfo{author}{\bibfnamefont{A.~A.} \bibnamefont{Kornyshev}},
  \bibnamefont{and} \bibinfo{author}{\bibfnamefont{M.~A.}
  \bibnamefont{Vorotyntsev}}, \bibinfo{journal}{Faraday Trans. II}
  \textbf{\bibinfo{volume}{72}}, \bibinfo{pages}{361} (\bibinfo{year}{1976}).

\bibitem[{\citenamefont{Li and Kardar}(1992)}]{Li:92}
\bibinfo{author}{\bibfnamefont{H.}~\bibnamefont{Li}} \bibnamefont{and}
  \bibinfo{author}{\bibfnamefont{M.}~\bibnamefont{Kardar}},
  \bibinfo{journal}{Phys. Rev. A} \textbf{\bibinfo{volume}{46}},
  \bibinfo{pages}{6490} (\bibinfo{year}{1992}).

\bibitem[{\citenamefont{Matyushov}(2004{\natexlab{b}})}]{DMjcp1:04}
\bibinfo{author}{\bibfnamefont{D.~V.} \bibnamefont{Matyushov}},
  \bibinfo{journal}{J. Chem. Phys.} \textbf{\bibinfo{volume}{120}},
  \bibinfo{pages}{1375} (\bibinfo{year}{2004}{\natexlab{b}}).

\bibitem[{\citenamefont{Press et~al.}(1996)\citenamefont{Press, Teukolsky,
  Vetterling, and Flannery}}]{Fortran:96}
\bibinfo{author}{\bibfnamefont{W.~H.} \bibnamefont{Press}},
  \bibinfo{author}{\bibfnamefont{S.~A.} \bibnamefont{Teukolsky}},
  \bibinfo{author}{\bibfnamefont{W.~T.} \bibnamefont{Vetterling}},
  \bibnamefont{and} \bibinfo{author}{\bibfnamefont{B.~P.}
  \bibnamefont{Flannery}}, \emph{\bibinfo{title}{Numerical recipes in {F}ortran
  77: {T}he art of scientific computing}} (\bibinfo{publisher}{Cambridge
  University Press}, \bibinfo{address}{Cambridge}, \bibinfo{year}{1996}).

\bibitem[{\citenamefont{Solomon and Lowery}(1993)}]{Solomon:93}
\bibinfo{author}{\bibfnamefont{E.~I.} \bibnamefont{Solomon}} \bibnamefont{and}
  \bibinfo{author}{\bibfnamefont{M.~D.} \bibnamefont{Lowery}},
  \bibinfo{journal}{Science} \textbf{\bibinfo{volume}{259}},
  \bibinfo{pages}{1575} (\bibinfo{year}{1993}).

\bibitem[{\citenamefont{Ungar et~al.}(1997)\citenamefont{Ungar, Scherer, and
  Voth}}]{Ungar:97}
\bibinfo{author}{\bibfnamefont{L.~W.} \bibnamefont{Ungar}},
  \bibinfo{author}{\bibfnamefont{N.~F.} \bibnamefont{Scherer}},
  \bibnamefont{and} \bibinfo{author}{\bibfnamefont{G.~A.} \bibnamefont{Voth}},
  \bibinfo{journal}{Biophys. J.} \textbf{\bibinfo{volume}{72}},
  \bibinfo{pages}{5} (\bibinfo{year}{1997}).

\bibitem[{\citenamefont{Ullmann et~al.}(1997)\citenamefont{Ullmann, Knapp, and
  Kosti{{\'c}}}}]{Ullmann:97}
\bibinfo{author}{\bibfnamefont{G.~M.} \bibnamefont{Ullmann}},
  \bibinfo{author}{\bibfnamefont{E.-W.} \bibnamefont{Knapp}}, \bibnamefont{and}
  \bibinfo{author}{\bibfnamefont{N.~M.} \bibnamefont{Kosti{{\'c}}}},
  \bibinfo{journal}{J. Am. Chem. Soc.} \textbf{\bibinfo{volume}{119}},
  \bibinfo{pages}{42} (\bibinfo{year}{1997}).

\bibitem[{\citenamefont{Comba and Remenyi}(2002)}]{Comba:02}
\bibinfo{author}{\bibfnamefont{P.}~\bibnamefont{Comba}} \bibnamefont{and}
  \bibinfo{author}{\bibfnamefont{R.}~\bibnamefont{Remenyi}},
  \bibinfo{journal}{J. Comput. Chem.} \textbf{\bibinfo{volume}{23}},
  \bibinfo{pages}{697} (\bibinfo{year}{2002}).

\bibitem[{\citenamefont{Werst et~al.}(1991)\citenamefont{Werst, Davoust, and
  Hoffman}}]{Werst:91}
\bibinfo{author}{\bibfnamefont{M.~M.} \bibnamefont{Werst}},
  \bibinfo{author}{\bibfnamefont{C.~E.} \bibnamefont{Davoust}},
  \bibnamefont{and} \bibinfo{author}{\bibfnamefont{B.~M.}
  \bibnamefont{Hoffman}}, \bibinfo{journal}{J. Am. Chem. Soc.}
  \textbf{\bibinfo{volume}{113}}, \bibinfo{pages}{1533} (\bibinfo{year}{1991}).

\bibitem[{\citenamefont{Libeu et~al.}(1997)\citenamefont{Libeu, Kikimoto,
  Nishiyama, Horinouchi, and Adman}}]{Libeu:97}
\bibinfo{author}{\bibfnamefont{C.~A.~P.} \bibnamefont{Libeu}},
  \bibinfo{author}{\bibfnamefont{M.}~\bibnamefont{Kikimoto}},
  \bibinfo{author}{\bibfnamefont{M.}~\bibnamefont{Nishiyama}},
  \bibinfo{author}{\bibfnamefont{S.}~\bibnamefont{Horinouchi}},
  \bibnamefont{and} \bibinfo{author}{\bibfnamefont{E.~T.} \bibnamefont{Adman}},
  \bibinfo{journal}{Biochemistry} \textbf{\bibinfo{volume}{36}},
  \bibinfo{pages}{13160} (\bibinfo{year}{1997}).

\bibitem[{\citenamefont{Kerpel and Ryde}(1999)}]{realq299}
\bibinfo{author}{\bibfnamefont{J.~O.~D.} \bibnamefont{Kerpel}}
  \bibnamefont{and} \bibinfo{author}{\bibfnamefont{U.}~\bibnamefont{Ryde}},
  \bibinfo{journal}{Proteins: Structure, Function, and Genetics}
  \textbf{\bibinfo{volume}{36}}, \bibinfo{pages}{157} (\bibinfo{year}{1999}).

\bibitem[{\citenamefont{Duan et~al.}(2003)\citenamefont{Duan, Wu, Chowdhury,
  Lee, Xiong, Zhang, Yang, Cieplak, Luo, Lee et~al.}}]{amberFF03}
\bibinfo{author}{\bibfnamefont{Y.}~\bibnamefont{Duan}},
  \bibinfo{author}{\bibfnamefont{C.}~\bibnamefont{Wu}},
  \bibinfo{author}{\bibfnamefont{S.}~\bibnamefont{Chowdhury}},
  \bibinfo{author}{\bibfnamefont{M.~C.} \bibnamefont{Lee}},
  \bibinfo{author}{\bibfnamefont{G.}~\bibnamefont{Xiong}},
  \bibinfo{author}{\bibfnamefont{W.}~\bibnamefont{Zhang}},
  \bibinfo{author}{\bibfnamefont{R.}~\bibnamefont{Yang}},
  \bibinfo{author}{\bibfnamefont{P.}~\bibnamefont{Cieplak}},
  \bibinfo{author}{\bibfnamefont{R.}~\bibnamefont{Luo}},
  \bibinfo{author}{\bibfnamefont{T.}~\bibnamefont{Lee}}, \bibnamefont{et~al.},
  \bibinfo{journal}{J.\ Comp.\ Chem.} \textbf{\bibinfo{volume}{24}},
  \bibinfo{pages}{1999} (\bibinfo{year}{2003}).

\bibitem[{\citenamefont{Cascella et~al.}(2006)\citenamefont{Cascella,
  Magistrato, Tavernelli, Carloni, and Rothlisberger}}]{Cascella:06}
\bibinfo{author}{\bibfnamefont{M.}~\bibnamefont{Cascella}},
  \bibinfo{author}{\bibfnamefont{A.}~\bibnamefont{Magistrato}},
  \bibinfo{author}{\bibfnamefont{I.}~\bibnamefont{Tavernelli}},
  \bibinfo{author}{\bibfnamefont{P.}~\bibnamefont{Carloni}}, \bibnamefont{and}
  \bibinfo{author}{\bibfnamefont{U.}~\bibnamefont{Rothlisberger}},
  \bibinfo{journal}{Proc. Natl. Acad. Sci.} \textbf{\bibinfo{volume}{103}},
  \bibinfo{pages}{19641} (\bibinfo{year}{2006}).

\bibitem[{\citenamefont{Blumberger and Klein}(2006)}]{Blumberger:06}
\bibinfo{author}{\bibfnamefont{J.}~\bibnamefont{Blumberger}} \bibnamefont{and}
  \bibinfo{author}{\bibfnamefont{M.~L.} \bibnamefont{Klein}},
  \bibinfo{journal}{J. Am. Chem. Soc.} \textbf{\bibinfo{volume}{128}},
  \bibinfo{pages}{13854} (\bibinfo{year}{2006}).

\bibitem[{\citenamefont{Guillot}(2002)}]{Guillot:02}
\bibinfo{author}{\bibfnamefont{B.}~\bibnamefont{Guillot}}, \bibinfo{journal}{J.
  Mol. Liq.} \textbf{\bibinfo{volume}{101}}, \bibinfo{pages}{219}
  (\bibinfo{year}{2002}).

\bibitem[{\citenamefont{Jorgensen et~al.}(1983)\citenamefont{Jorgensen,
  Chandrasekhar, Madura, Impey, and Klein}}]{tip3p:83}
\bibinfo{author}{\bibfnamefont{W.~L.} \bibnamefont{Jorgensen}},
  \bibinfo{author}{\bibfnamefont{J.}~\bibnamefont{Chandrasekhar}},
  \bibinfo{author}{\bibfnamefont{J.~D.} \bibnamefont{Madura}},
  \bibinfo{author}{\bibfnamefont{R.~W.} \bibnamefont{Impey}}, \bibnamefont{and}
  \bibinfo{author}{\bibfnamefont{M.~L.} \bibnamefont{Klein}},
  \bibinfo{journal}{J. Chem. Phys.} \textbf{\bibinfo{volume}{79}},
  \bibinfo{pages}{926} (\bibinfo{year}{1983}).

\bibitem[{\citenamefont{Wertheim}(1971)}]{Wertheim:71}
\bibinfo{author}{\bibfnamefont{M.~S.} \bibnamefont{Wertheim}},
  \bibinfo{journal}{J. Chem. Phys.} \textbf{\bibinfo{volume}{55}},
  \bibinfo{pages}{4291} (\bibinfo{year}{1971}).

\bibitem[{\citenamefont{Wertheim}(1979)}]{Wertheim:79}
\bibinfo{author}{\bibfnamefont{M.~S.} \bibnamefont{Wertheim}},
  \bibinfo{journal}{Molec. Phys.} \textbf{\bibinfo{volume}{37}},
  \bibinfo{pages}{83} (\bibinfo{year}{1979}).

\bibitem[{\citenamefont{Gupta and Matyushov}(2004)}]{DMjpca:04}
\bibinfo{author}{\bibfnamefont{S.}~\bibnamefont{Gupta}} \bibnamefont{and}
  \bibinfo{author}{\bibfnamefont{D.~V.} \bibnamefont{Matyushov}},
  \bibinfo{journal}{J. Phys. Chem. A} \textbf{\bibinfo{volume}{108}},
  \bibinfo{pages}{2087} (\bibinfo{year}{2004}).

\bibitem[{\citenamefont{Case et~al.}(2005)\citenamefont{Case, III, Darden,
  Gohlke, Luo, Jr., Onufriev, Simmerling, Wang, and Woods}}]{amber8}
\bibinfo{author}{\bibfnamefont{D.~A.} \bibnamefont{Case}},
  \bibinfo{author}{\bibfnamefont{T.~E.~C.} \bibnamefont{III}},
  \bibinfo{author}{\bibfnamefont{T.}~\bibnamefont{Darden}},
  \bibinfo{author}{\bibfnamefont{H.}~\bibnamefont{Gohlke}},
  \bibinfo{author}{\bibfnamefont{R.}~\bibnamefont{Luo}},
  \bibinfo{author}{\bibfnamefont{K.~M.~M.} \bibnamefont{Jr.}},
  \bibinfo{author}{\bibfnamefont{A.}~\bibnamefont{Onufriev}},
  \bibinfo{author}{\bibfnamefont{C.}~\bibnamefont{Simmerling}},
  \bibinfo{author}{\bibfnamefont{B.}~\bibnamefont{Wang}}, \bibnamefont{and}
  \bibinfo{author}{\bibfnamefont{R.~J.} \bibnamefont{Woods}},
  \bibinfo{journal}{J.\ Comp.\ Chem.} \textbf{\bibinfo{volume}{26}},
  \bibinfo{pages}{1668} (\bibinfo{year}{2005}).

\bibitem[{\citenamefont{Xue et~al.}(1998)\citenamefont{Xue, Okvist, Hansson,
  and Young}}]{1ag698}
\bibinfo{author}{\bibfnamefont{Y.}~\bibnamefont{Xue}},
  \bibinfo{author}{\bibfnamefont{M.}~\bibnamefont{Okvist}},
  \bibinfo{author}{\bibfnamefont{O.}~\bibnamefont{Hansson}}, \bibnamefont{and}
  \bibinfo{author}{\bibfnamefont{S.}~\bibnamefont{Young}},
  \bibinfo{journal}{Prot. Sci.} \textbf{\bibinfo{volume}{7}},
  \bibinfo{pages}{2099} (\bibinfo{year}{1998}).

\bibitem[{\citenamefont{Berendsen et~al.}(1984)\citenamefont{Berendsen, Postma,
  van Gunsteren, DiNola, and Haak}}]{berend84}
\bibinfo{author}{\bibfnamefont{H.~J.~C.} \bibnamefont{Berendsen}},
  \bibinfo{author}{\bibfnamefont{J.~P.~M.} \bibnamefont{Postma}},
  \bibinfo{author}{\bibfnamefont{W.~F.} \bibnamefont{van Gunsteren}},
  \bibinfo{author}{\bibfnamefont{A.}~\bibnamefont{DiNola}}, \bibnamefont{and}
  \bibinfo{author}{\bibfnamefont{J.~R.} \bibnamefont{Haak}},
  \bibinfo{journal}{J.\ Chem.\ Phys.} \textbf{\bibinfo{volume}{81}},
  \bibinfo{pages}{3684} (\bibinfo{year}{1984}).

\bibitem[{\citenamefont{Reiss}(1988)}]{Reiss:88}
\bibinfo{author}{\bibfnamefont{H.}~\bibnamefont{Reiss}}, \bibinfo{journal}{J.\
  Electrochem.\ Soc.} \textbf{\bibinfo{volume}{135}}, \bibinfo{pages}{247C}
  (\bibinfo{year}{1988}).

\bibitem[{\citenamefont{Landau and Lifshits}(1980)}]{Landau5}
\bibinfo{author}{\bibfnamefont{L.~D.} \bibnamefont{Landau}} \bibnamefont{and}
  \bibinfo{author}{\bibfnamefont{E.~M.} \bibnamefont{Lifshits}},
  \emph{\bibinfo{title}{Statistical Physics}} (\bibinfo{publisher}{Pergamon
  Press}, \bibinfo{address}{New York}, \bibinfo{year}{1980}).

\bibitem[{\citenamefont{Bockris}(1970)}]{Bockris:70}
\bibinfo{author}{\bibfnamefont{J.~O.} \bibnamefont{Bockris}},
  \emph{\bibinfo{title}{Modern Electrochemistry}}
  (\bibinfo{publisher}{McDonald}, \bibinfo{address}{London},
  \bibinfo{year}{1970}).

\bibitem[{\citenamefont{Schmickler}(1996)}]{Schmickler:96}
\bibinfo{author}{\bibfnamefont{W.}~\bibnamefont{Schmickler}},
  \emph{\bibinfo{title}{Interfacial Electrochemistry}}
  (\bibinfo{publisher}{Oxford University Press}, \bibinfo{address}{New York},
  \bibinfo{year}{1996}).

\bibitem[{\citenamefont{Marcus}(1965)}]{Marcus:65}
\bibinfo{author}{\bibfnamefont{R.~A.} \bibnamefont{Marcus}},
  \bibinfo{journal}{J. Chem. Phys.} \textbf{\bibinfo{volume}{43}},
  \bibinfo{pages}{679} (\bibinfo{year}{1965}).

\bibitem[{\citenamefont{Reiss}(1985)}]{Reiss:85}
\bibinfo{author}{\bibfnamefont{H.}~\bibnamefont{Reiss}}, \bibinfo{journal}{J.
  Phys. Chem.} \textbf{\bibinfo{volume}{89}}, \bibinfo{pages}{3783}
  (\bibinfo{year}{1985}).

\bibitem[{\citenamefont{Gorodyskii et~al.}(1991)\citenamefont{Gorodyskii,
  Karasevskii, and Matyushov}}]{DMjec:91}
\bibinfo{author}{\bibfnamefont{A.~V.} \bibnamefont{Gorodyskii}},
  \bibinfo{author}{\bibfnamefont{A.~I.} \bibnamefont{Karasevskii}},
  \bibnamefont{and} \bibinfo{author}{\bibfnamefont{D.~V.}
  \bibnamefont{Matyushov}}, \bibinfo{journal}{J. Electroanal. Chem.}
  \textbf{\bibinfo{volume}{315}}, \bibinfo{pages}{9} (\bibinfo{year}{1991}).

\bibitem[{\citenamefont{Swartz and Ichiye}(1996)}]{Swartz:96}
\bibinfo{author}{\bibfnamefont{P.~D.} \bibnamefont{Swartz}} \bibnamefont{and}
  \bibinfo{author}{\bibfnamefont{T.}~\bibnamefont{Ichiye}},
  \bibinfo{journal}{Biochemistry} \textbf{\bibinfo{volume}{35}},
  \bibinfo{pages}{13772} (\bibinfo{year}{1996}).

\bibitem[{\citenamefont{Moore et~al.}(1986)\citenamefont{Moore, Pettigrew, and
  Rogers}}]{Moore:86}
\bibinfo{author}{\bibfnamefont{G.~R.} \bibnamefont{Moore}},
  \bibinfo{author}{\bibfnamefont{G.~W.} \bibnamefont{Pettigrew}},
  \bibnamefont{and} \bibinfo{author}{\bibfnamefont{N.~K.}
  \bibnamefont{Rogers}}, \bibinfo{journal}{Proc. Natl. Acad. Sci. USA}
  \textbf{\bibinfo{volume}{83}}, \bibinfo{pages}{4998} (\bibinfo{year}{1986}).

\bibitem[{\citenamefont{Hotta et~al.}(2005)\citenamefont{Hotta, Kimura, and
  Sasai}}]{Hotta:05}
\bibinfo{author}{\bibfnamefont{T.}~\bibnamefont{Hotta}},
  \bibinfo{author}{\bibfnamefont{A.}~\bibnamefont{Kimura}}, \bibnamefont{and}
  \bibinfo{author}{\bibfnamefont{M.}~\bibnamefont{Sasai}}, \bibinfo{journal}{J.
  Phys. Chem. B} \textbf{\bibinfo{volume}{109}}, \bibinfo{pages}{18600}
  (\bibinfo{year}{2005}).

\bibitem[{\citenamefont{Stillinger}(1973)}]{Stillinger:73}
\bibinfo{author}{\bibfnamefont{F.~H.} \bibnamefont{Stillinger}},
  \bibinfo{journal}{J.\ Solut.\ Chem.} \textbf{\bibinfo{volume}{2}},
  \bibinfo{pages}{141} (\bibinfo{year}{1973}).

\bibitem[{\citenamefont{Hummer and Garde}(1998)}]{HummerPRL:98}
\bibinfo{author}{\bibfnamefont{G.}~\bibnamefont{Hummer}} \bibnamefont{and}
  \bibinfo{author}{\bibfnamefont{S.}~\bibnamefont{Garde}},
  \bibinfo{journal}{Phys. Rev. Lett.} \textbf{\bibinfo{volume}{80}},
  \bibinfo{pages}{4193} (\bibinfo{year}{1998}).

\bibitem[{\citenamefont{Chandler}(2005)}]{ChandlerNature:05}
\bibinfo{author}{\bibfnamefont{D.}~\bibnamefont{Chandler}},
  \bibinfo{journal}{Nature} \textbf{\bibinfo{volume}{437}},
  \bibinfo{pages}{640} (\bibinfo{year}{2005}).

\bibitem[{\citenamefont{Florian and Warshel}(1999)}]{Florian:99}
\bibinfo{author}{\bibfnamefont{J.}~\bibnamefont{Florian}} \bibnamefont{and}
  \bibinfo{author}{\bibfnamefont{A.}~\bibnamefont{Warshel}},
  \bibinfo{journal}{J. Phys. Chem. B} \textbf{\bibinfo{volume}{103}},
  \bibinfo{pages}{10282} (\bibinfo{year}{1999}).

\bibitem[{\citenamefont{Ghorai and Matyushov}(2006{\natexlab{b}})}]{DMjpcb1:06}
\bibinfo{author}{\bibfnamefont{P.~K.} \bibnamefont{Ghorai}} \bibnamefont{and}
  \bibinfo{author}{\bibfnamefont{D.~V.} \bibnamefont{Matyushov}},
  \bibinfo{journal}{J. Phys. Chem. B} \textbf{\bibinfo{volume}{110}},
  \bibinfo{pages}{1866} (\bibinfo{year}{2006}{\natexlab{b}}).

\bibitem[{\citenamefont{Ashbaugh}(2000)}]{Ashbaugh:00}
\bibinfo{author}{\bibfnamefont{H.~S.} \bibnamefont{Ashbaugh}},
  \bibinfo{journal}{J. Phys. Chem. B} \textbf{\bibinfo{volume}{104}},
  \bibinfo{pages}{7235} (\bibinfo{year}{2000}).

\bibitem[{\citenamefont{Rajamani et~al.}(2004)\citenamefont{Rajamani, Ghosh,
  and Garde}}]{Rajamani:04}
\bibinfo{author}{\bibfnamefont{S.}~\bibnamefont{Rajamani}},
  \bibinfo{author}{\bibfnamefont{T.}~\bibnamefont{Ghosh}}, \bibnamefont{and}
  \bibinfo{author}{\bibfnamefont{S.}~\bibnamefont{Garde}},
  \bibinfo{journal}{J.\ Chem.\ Phys.} \textbf{\bibinfo{volume}{120}},
  \bibinfo{pages}{4457} (\bibinfo{year}{2004}).

\bibitem[{\citenamefont{Cerutti et~al.}(2007)\citenamefont{Cerutti, Baker, and
  McCammon}}]{Cerutti:07}
\bibinfo{author}{\bibfnamefont{D.~S.} \bibnamefont{Cerutti}},
  \bibinfo{author}{\bibfnamefont{N.~A.} \bibnamefont{Baker}}, \bibnamefont{and}
  \bibinfo{author}{\bibfnamefont{J.~A.} \bibnamefont{McCammon}},
  \bibinfo{journal}{J.\ Chem.\ Phys.} \textbf{\bibinfo{volume}{127}},
  \bibinfo{eid}{155101} (\bibinfo{year}{2007}).

\bibitem[{\citenamefont{Pratt and Chandler}(1977)}]{PrattC:77}
\bibinfo{author}{\bibfnamefont{L.~R.} \bibnamefont{Pratt}} \bibnamefont{and}
  \bibinfo{author}{\bibfnamefont{D.}~\bibnamefont{Chandler}},
  \bibinfo{journal}{J. Chem. Phys.} \textbf{\bibinfo{volume}{67}},
  \bibinfo{pages}{3683} (\bibinfo{year}{1977}).

\bibitem[{\citenamefont{Garde et~al.}(1996)\citenamefont{Garde, Hummer,
  Garc\'{\i}a, Paulaitis, and Pratt}}]{GardePRL:96}
\bibinfo{author}{\bibfnamefont{S.}~\bibnamefont{Garde}},
  \bibinfo{author}{\bibfnamefont{G.}~\bibnamefont{Hummer}},
  \bibinfo{author}{\bibfnamefont{A.~E.} \bibnamefont{Garc\'{\i}a}},
  \bibinfo{author}{\bibfnamefont{M.~E.} \bibnamefont{Paulaitis}},
  \bibnamefont{and} \bibinfo{author}{\bibfnamefont{L.~R.} \bibnamefont{Pratt}},
  \bibinfo{journal}{Phys.\ Rev.\ Lett.} \textbf{\bibinfo{volume}{77}},
  \bibinfo{pages}{4966} (\bibinfo{year}{1996}).

\bibitem[{\citenamefont{Ashbaugh and Pratt}(2006)}]{Ashbaugh:06}
\bibinfo{author}{\bibfnamefont{H.~S.} \bibnamefont{Ashbaugh}} \bibnamefont{and}
  \bibinfo{author}{\bibfnamefont{L.~R.} \bibnamefont{Pratt}},
  \bibinfo{journal}{Rev.\ Mod.\ Phys.} \textbf{\bibinfo{volume}{78}},
  \bibinfo{eid}{159} (\bibinfo{year}{2006}).

\bibitem[{\citenamefont{Lum et~al.}(1999)\citenamefont{Lum, Chandler, and
  Weeks}}]{Lum:99}
\bibinfo{author}{\bibfnamefont{K.}~\bibnamefont{Lum}},
  \bibinfo{author}{\bibfnamefont{D.}~\bibnamefont{Chandler}}, \bibnamefont{and}
  \bibinfo{author}{\bibfnamefont{J.}~\bibnamefont{Weeks}},
  \bibinfo{journal}{J.\ Phys.\ Chem.\ B} \textbf{\bibinfo{volume}{103}},
  \bibinfo{pages}{4570} (\bibinfo{year}{1999}).

\bibitem[{\citenamefont{Huang and Chandler}(2000)}]{Huang:00}
\bibinfo{author}{\bibfnamefont{D.~M.} \bibnamefont{Huang}} \bibnamefont{and}
  \bibinfo{author}{\bibfnamefont{D.}~\bibnamefont{Chandler}},
  \bibinfo{journal}{Phys. Rev. E} \textbf{\bibinfo{volume}{61}},
  \bibinfo{pages}{1501} (\bibinfo{year}{2000}).

\bibitem[{\citenamefont{Fenimore et~al.}(2004)\citenamefont{Fenimore,
  Frauenfelder, McMahon, and Young}}]{Fenimore:04}
\bibinfo{author}{\bibfnamefont{P.~W.} \bibnamefont{Fenimore}},
  \bibinfo{author}{\bibfnamefont{H.}~\bibnamefont{Frauenfelder}},
  \bibinfo{author}{\bibfnamefont{B.~H.} \bibnamefont{McMahon}},
  \bibnamefont{and} \bibinfo{author}{\bibfnamefont{R.~D.} \bibnamefont{Young}},
  \bibinfo{journal}{Proc. Natl. Acad. Sci.} \textbf{\bibinfo{volume}{101}},
  \bibinfo{pages}{14408} (\bibinfo{year}{2004}).

\bibitem[{\citenamefont{Dzubiella et~al.}(2006)\citenamefont{Dzubiella,
  Swanson, and McCammon}}]{Dzubiella:06}
\bibinfo{author}{\bibfnamefont{J.}~\bibnamefont{Dzubiella}},
  \bibinfo{author}{\bibfnamefont{J.~M.~J.} \bibnamefont{Swanson}},
  \bibnamefont{and} \bibinfo{author}{\bibfnamefont{J.~A.}
  \bibnamefont{McCammon}}, \bibinfo{journal}{Phys.\ Rev.\ Lett.}
  \textbf{\bibinfo{volume}{96}}, \bibinfo{eid}{087802} (\bibinfo{year}{2006}).

\bibitem[{\citenamefont{Hupp et~al.}(1993)\citenamefont{Hupp, Dong, Blackbourn,
  and Lu}}]{Hupp:93}
\bibinfo{author}{\bibfnamefont{J.~T.} \bibnamefont{Hupp}},
  \bibinfo{author}{\bibfnamefont{Y.}~\bibnamefont{Dong}},
  \bibinfo{author}{\bibfnamefont{R.~L.} \bibnamefont{Blackbourn}},
  \bibnamefont{and} \bibinfo{author}{\bibfnamefont{H.}~\bibnamefont{Lu}},
  \bibinfo{journal}{J. Phys. Chem.} \textbf{\bibinfo{volume}{97}},
  \bibinfo{pages}{3278} (\bibinfo{year}{1993}).

\bibitem[{\citenamefont{D'Alessandro et~al.}(2005)\citenamefont{D'Alessandro,
  Junk, and Keene}}]{Dalessandro:05}
\bibinfo{author}{\bibfnamefont{D.~M.} \bibnamefont{D'Alessandro}},
  \bibinfo{author}{\bibfnamefont{P.~C.} \bibnamefont{Junk}}, \bibnamefont{and}
  \bibinfo{author}{\bibfnamefont{F.~R.} \bibnamefont{Keene}},
  \bibinfo{journal}{Supramolecular Chemistry} \textbf{\bibinfo{volume}{17}},
  \bibinfo{pages}{529} (\bibinfo{year}{2005}).

\end{thebibliography}

\end{document}